\documentclass{article}
\pdfoutput=1

\usepackage{arxiv}
\usepackage[utf8]{inputenc} % allow utf-8 input
\usepackage[T1]{fontenc}    % use 8-bit T1 fonts
\usepackage[hyphens]{url}   % simple URL typesetting
\usepackage{hyperref}       % hyperlinks
\usepackage{booktabs}       % professional-quality tables
\usepackage{amsfonts}       % blackboard math symbols
\usepackage{nicefrac}       % compact symbols for 1/2, etc.
\usepackage{microtype}      % microtypography
\usepackage{lipsum}

\usepackage{amsmath}
\usepackage{graphicx}
\usepackage{footnote}
\usepackage{footmisc} % for footnote referencing
\usepackage{tablefootnote}
\makesavenoteenv{tabular}	% Add to show footnotes also in tables
\makesavenoteenv{table}
\usepackage{multirow}
\usepackage{units}
\usepackage{enumitem} % change enumeration sign
\usepackage{gensymb} % for the degree sign
\usepackage{xcolor,colortbl}
\definecolor{lightgray}{gray}{0.95}
\usepackage{makecell} % for line break in cell
\usepackage{authblk}

\title{A Survey of Data Quality Measurement and Monitoring Tools}

\author[1, 2]{\bf Lisa Ehrlinger}
\author[1]{\bf Elisa Rusz}
\author[1]{\bf Wolfram W{\"o}{\ss}}
\affil[1]{\footnotesize Johannes Kepler University Linz, Austria}
\affil[ ]{\texttt{lisa.ehrlinger@jku.at}, \texttt{elisa.rusz@jku.at}, \texttt{wolfram.woess@jku.at}}
\affil[2]{\footnotesize Software Competence Center Hagenberg, Austria}
\affil[ ]{\texttt{lisa.ehrlinger@scch.at}}

\begin{document}
\maketitle

\begin{abstract}
High-quality data is key to interpretable and trustworthy data analytics and the basis for meaningful data-driven decisions. 
In practical scenarios, data quality is typically associated with data preprocessing, profiling, and cleansing for subsequent tasks like data integration or data analytics. 
However, from a scientific perspective, a lot of research has been published about the measurement (i.e., the detection) of data quality issues and different generally applicable data quality dimensions and metrics have been discussed. 
In this work, we close the gap between research into data quality measurement and practical implementations by investigating the functional scope of current data quality tools. 
With a systematic search, we identified 667 software tools dedicated to ``data quality'', from which we evaluated 13 tools with respect to three functionality areas: (1) data profiling, (2) data quality measurement in terms of metrics, and (3) continuous data quality monitoring. We selected the evaluated tools with regard to pre-defined exclusion criteria to ensure that they are domain-independent, provide the investigated functions, and are evaluable freely or as trial. 
This survey aims at a comprehensive overview on state-of-the-art data quality tools and reveals potential for their functional enhancement. Additionally, the results allow a critical discussion on concepts, which are widely accepted in research, but hardly implemented in any tool observed, for example, generally applicable data quality metrics. 
\end{abstract}

% keywords can be removed

\section{Introduction}
% General motivation for DQ
% essential 
Data quality (DQ) measurement is a fundamental building block for estimating the relevance of data-driven decisions. 
Such decisions accompany our everyday life, for instance, machine-based decision in ranking algorithms, industrial robots, and self-driving cars in the emerging field of artificial intelligence. Also human-based decisions rely on high-quality data, for example, the decision whether to promote or to suspend the production of a specific product is usually based on sales data. Despite the clear correlation between data and decision quality, 84~\% of the CEOs in the US are concerned about their DQ~\cite{KPMG_2016} and ``organizations believe poor data quality to be responsible for an average of \$15 million per year in losses''~\cite{Moore_2018}.
%Thus, DQ is no longer a question of ``hygiene'', but has become critical for operational excellence and is perceived as the greatest challenge in corporate data management~\cite{OttoOesterle_2016}. 
Thus, DQ is ``no longer a question of `hygiene' [...], but rather has become critical for operational excellence''~\cite{OttoOesterle_2016} and is perceived as the greatest challenge in corporate data management (cf.~\cite{OttoOesterle_2016}).
% 83 % of the respondents of a survey conducted by Experian Information  Solutions[2016]state  that poor  data  quality  has actually  hurt  their  business  objectives,  and  66% report  that  poor  data  quality  has  had  a  negative  impact  on  their  organization  in  the  last  twelve  months.

To increase the trust in data-driven decisions, it is necessary to measure and to know the quality of the employed data with appropriate tools~\cite{QuaIIeJournal_2018, Heinrich_2018}. DQ measurement is a prerequisite to comprehensive and strategic DQ improvement (i.e., data cleansing). Yet, according to a German survey, 66~\% of companies use Excel or Access solutions to validate their DQ and 63~\% of the companies determine their DQ manually and ad hoc without any long-term DQ management strategy~\cite{Schaeffer_2014}. However, comprehensive DQ management requires a defined strategy, which is carried out continuously~\cite{Coleman_2012}, as well as appropriate tools to implement and automate such a strategy~\cite{EhrlingerWoess_2017}. 
% Although most existing DQ methodologies describe DQ management as a cyclic process (cf.~\cite{Wang_1998,Lee_2009Journey,English_1999,Redman_1997}), DQ tools usually allow to perform DQ tasks at a given point in time, which can be repeated subsequently, but they do not offer a comprehensive solution to track the evolvement of specific DQ measurements. 
%TODO: cite methodologies

%% Specify the specific problem we are tackling: practice vs. theory,
Research about data quality has been conducted since the 1980s and since then, DQ is most often associated with the ``fitness for use'' principle~\cite{WandWang_1996,WangStrong_1996,Chrisman_1984}, which refers to the subjectivity and context-dependency of this topic. Data quality is typically referred to as a multi-dimensional concept, where single aspects are described by DQ \textit{dimensions} (e.g., accuracy, completeness, timeliness). The fulfillment of a DQ dimension can be quantified using one or several DQ \textit{metrics}~\cite{QuaIIeJournal_2018}. According to~\cite{IEEEStandard_QualityMetric}, a metric is a formula that yields a numerical value. 
In parallel to the scientific background, a wide variety of commercial, open-source, and academic DQ tools with different foci have been developed since then. The range of functions offered by those tools varies widely, because the term ``data quality'' is context-dependent and not always used consistently. 
Despite the large number of publications, tools, and concepts into data quality, it is not always clear how to map the concepts from the theory (i.e., dimensions and metrics) to a practical implementation (i.e. tools). Therefore, the question of how to measure and monitor DQ is still not sufficiently answered~\cite{Coleman_2012}.
% Our contribution
The main contribution of this survey is a comprehensive overview on state-of-the-art DQ tools with respect to their DQ measurement and monitoring functionalities in order to answer the question \textit{how data quality should actually be measured and monitored automatically. }

%%% Summary; what we actually do in the survey
To tackle this question in a scientific way, we conducted a systematic search, where we identified 667 software tools dedicated to ``data quality''. According to predefined exclusion criteria, we selected 13 DQ tools for deeper investigation. To systematically evaluate the functional scope of the tools, we developed a requirements catalog comprising three categories: (1) data profiling, (2) DQ measurement in terms of dimension and metrics, and (3) continuous DQ monitoring. 
While some tools, which we found in our search, solely offer data cleansing and improvement functionality, we specifically observe the measurement capabilities, that is, the detection of DQ issues. Since an automated modification of the data (i.e., data cleansing) is usually not possible in productive information systems with critical content, tools that detect and report DQ issues are required. 
Additionally, we believe that ongoing monitoring is required to ensure high-quality data over time. 

% Main findings
The results of this survey are not only relevant for DQ professionals to select the most appropriate tool for a given use case, but also  highlight the current capabilities of state-of-the-art DQ tools. Especially since such a wide variety of DQ tools exist, it is often not clear which functional scope can be expected. The main findings of this article can be summarized as follows: 
\begin{itemize}
    \item Despite the presumption that the emerging market of DQ tools is still under development (cf.~\cite{GartnerDQTools2017}), we found a vast number (667) of DQ tools through our systematic search, where most of them have never been included in one of the existing surveys.
    \item Approximately half (50.82~\%) of the DQ tools were domain specific, which means they were either dedicated to specific types of data or built to measure the DQ of a proprietary tool. 
    \item 16.67~\% of the DQ tools focused on data cleansing without a proper DQ measurement strategy. 
    \item Most surveyed tools supported data profiling to some extent, but considering the research state, there is potential for functional enhancement in data profiling, especially with respect to multi-column profiling and dependency discovery. 
     \item We did not find a tool that implements a wider range of DQ metrics for the most important DQ dimensions as proposed in research papers (cf.~\cite{Heinrich_2018,Piro_2014,BatiniScannapieco_2016}). Identified metric implementations have several drawbacks: some are only applicable on attribute-level (e.g., no aggregation), some require a gold standard that might not exist, and some have implementation errors.  
     \item In general-purpose DQ tools, DQ monitoring is considered a premium feature, which is liable to costs and only provided in professional versions. Exceptions are dedicated open-source DQ monitoring tools, like Apache Griffin or MobyDQ, which support the automation of rules, but lack pre-defined functions and data profiling capabilities.
\end{itemize}

% Structure of this article
This article is structured as follows: Section~\ref{sec:methodology} covers the applied methodology to conduct this research, including related surveys, our research questions, and the tool selection strategy. Section~\ref{sec:theory} gives an overview on data quality, its measurement and monitoring, and represents our evaluation framework and requirements catalog. In Section~\ref{sec:eval}, we describe the tools, which have been selected for investigation, and discuss the evaluation. 
The results and lessons learned are summarized in Section~\ref{sec:survDisc}.
We conclude in Section~\ref{sec:conclusion} with an outlook on future work. 
\section{Survey Methodology}
\label{sec:methodology}
A systematic survey is usually started by defining a ``protocol that specifices the research questions being addressed and the methods that will be used''~\cite{Kitchenham_2004}. 
This section describes the protocol we developed to systematically conduct our survey. The structure of the protocol has been derived from the methodology for systematic reviews in computer science by Kitchenham~\cite{Kitchenham_2004}. 
Since the focus in~\cite{Kitchenham_2004} is on the evaluation of primary research papers and not on specific implementations, we omit steps 5, 6, and 7 of the suggested planning information, including quality assessment, a data extraction strategy, and the synthesis of the extracted data from the original research papers.

\subsection{Related Surveys}
\label{sec:relSurveys}
Although a lot of DQ methods and tools have been published, there are few scientific studies about the functional scope of DQ tools. 
Gartner Inc.~\cite{GartnerDQTools2016,GartnerDQTools2017,GartnerDQTools2019} lists the strengths and cautions of vendors of commercial DQ tools in their ``Magic Quadrant for Data Quality Tools'' 2016 (17 vendors), 2017 (16 vendors), and 2019 (15 vendors). 
They include vendors that offer software tools or cloud-based services, which deliver general-purpose DQ functionalities, including at least profiling, parsing, standardization/cleansing, matching, and monitoring~\cite{GartnerDQTools2017}.
The study is vendor-focused and does not provide a detailed comparison of the respective data quality tools in terms of functionality (e.g., measurement and monitoring capabilities). 
However, the ``Magic Quadrant for Data Quality Tools'' contains a representative selection of commercial DQ tools, which is a valuable complement to our survey.
A detailed survey that compares 11 selected DQ tools in terms of algorithms and functionality has been published by Fraunhofer IAO~\cite{Fraunhofer2012} in 2012. While this study is probably the closest to our work from the structure of the tool comparison,~\cite{Fraunhofer2012} focuses on tools popular in Germany and is only available in German. In addition, we aim at a scientific approach to observe the availability of DQ tools from a general perspective by also justifying the tool selection. 

Woodall et al.~\cite{Woodall_2014} aim at a classification of DQ assessment and improvement methods. They understand DQ methods as automatically executable algorithms to detect and/or correct DQ problems, for example, column analysis, data verification, or data standardization. 
As basis for their classification, they reviewed the list of DQ tools included in the ``Magic Quadrant for Data Quality Tools 2012'' by Gartner and extracted a list of offered DQ methods that tackle specific DQ problems. Woodall et al. do not provide an in-depth comparison of which method is contained in which tool since their focus is on the method classification. 

Barateiro and Galhardas~\cite{Barateiro_DQSurvey_2005} compared 9 academic and 28 commercial DQ tools in a 2005 scientific survey. The paper does not cover state-of-the-art tools and the survey was not conducted in a systematic way, which means, it is unclear how the list of DQ tools has been selected. 
In addition, the authors state that DQ tools aim at detecting and correcting data problems, which is why they observe functionalities for both, DQ measurement as well as data cleansing, with an emphasis on the second aspect. 
In contrast, we focus on the measurement of data quality issues only, with special consideration of long-term monitoring functionality. 

In 2010, Pushkarev et al.~\cite{Pushkarev_DQSurvey_2010} proposed an overview of 7 open-source or freely available DQ tools. They described each tool briefly and compared the functionalities of the tools by means of \textit{performance criteria} (including 6 usability features like data source connectivity, report creation, or the graphical user interface -- GUI) and \textit{core functionality}. The core functionality consists of 4 groups, which are further subdivided into specific features that are observed: data profiling (e.g., data pattern discovery), data integration (e.g., ETL), data cleansing (e.g., parsing and standardization), and data monitoring. 
Due to the limited number of pages, Pushkarev et al. do not provide detailed insights in the implementation of specific criteria, and mainly distinguish between the availability of a feature (Y) or its absence (N). 
For example, the authors list 9 usability criteria for the GUI, but in the evaluation they only distinguish between (g) representing ``not user friendly GUI'' and a (G) for ``user-friendly GUI'' with drag and drop functionality. 
Pulla et al.~\cite{Pulla_DQSurvey_2016} published a revised version of the tool overview in 2016, which is very similar to the original work in terms of structure and methodology. They used the same criteria structure as Pushkarev et al.~\cite{Pushkarev_DQSurvey_2010}, but omitted the data monitoring group (since it is not provided by any of the tools~\cite{Pulla_DQSurvey_2016}) and 4 other sub-features without further justification. The list of investigated DQ tools was extended from 7 to 10. 
Our survey differs notably from those two papers since we aim at a systematic and documented approach to select DQ tools, while~\cite{Pushkarev_DQSurvey_2010} and~\cite{Pulla_DQSurvey_2016} presented a pre-defined selection of free or open-source tools without publishing their selection strategy. 
Moreover, we do not consider tools that have been primarily developed for a specific data management task (e.g., data integration or data cleansing) but focus on data profiling, DQ measurement, and DQ monitoring. We evaluate these feature groups by means of a more detailed and comprehensive criteria catalog as provided by other published surveys mentioned above. 

Another study by Gao et al.~\cite{GaoEtAl_BigDataQA_2016},  published in 2016, focuses on big data quality assurance. 
However, the authors did not clarify the methodology, that is, the selection of the investigated tools and evaluation criteria. 
In contrast to our survey, were the focus is on the actual DQ measurement functionalities, the comparison in~\cite{GaoEtAl_BigDataQA_2016} includes mainly technical features like the supported operating system and data sources, as well as a limited list of 4 basic data validation functions. 

To the best of our knowledge there exists no systematic survey that evaluates state-of-the-art tools with respect to DQ \textit{measurement and monitoring} in detail. In this paper, we address this topic.

\subsection{Research Questions}
The aim of this survey is to evaluate and compare existing DQ tools with respect to their DQ measurement and monitoring functionalities in order to answer the general research question \textit{how DQ measurement and monitoring is implemented in state-of-the-art DQ tools.} This question can be refined with three more specific research questions, where the theoretical background is discussed in more detail in Section~\ref{sec:theory}. For a detailed evaluation, we present a requirements catalog in Section~\ref{sec:req_catalog} that maps our research questions to concrete technical requirements. 

\begin{enumerate}
\item Which data profiling capabilities are supported by current DQ tools?
\item Which data quality dimensions and metrics can be measured with current DQ tools?
\item Do DQ tools allow continuous data quality monitoring over time?
\end{enumerate}

\subsection{DQ Tool Search Strategy}
To establish a comprehensive list of existing DQ tools, we adopted a three-fold strategy.
First, we included all observed tools from previous surveys by  Barateiro and Gallhardas~\cite{Barateiro_DQSurvey_2005}, Fraunhofer IAO~\cite{Fraunhofer2012}, Gao et al.~\cite{GaoEtAl_BigDataQA_2016}, Gartner Inc.~\cite{GartnerDQTools2017}, Pulla et al.~\cite{Pulla_DQSurvey_2016}, and Pushkarev et al.~\cite{Pushkarev_DQSurvey_2010} as candidate tools. 
Second, we conducted a systematic search to find research papers that introduce academic and open-source DQ tools. 
The third part of our search strategy consists of a random Google search by using the same search term combinations as for the systematic search. 
In contrast to the systematic search, we do not aim at a comprehensive observation of all search results, which is unfeasible for Google search results. 
However, to also identify non-research tools that have not been described in scientific papers, we consider this random search as enrichment to guarantee a best possible coverage of candidate tools. 
The remainder of this section is dedicated to the systematic search. 

We identified the following search terms to conduct the systematic search: \textit{data quality}, \textit{information quality}, and \textit{tool}. 
Since ``information quality'' is considered a synonym to ``data quality''~\cite{ZhuLeeWang_2014}, we applied both search terms to achieve higher coverage. 
We decided not to add the terms ``assessment'' and ``monitoring'' to the search, as it would automatically exclude tools that do not specifically use these keywords. 
Consequently, the following search expression has been applied: 

\begin{quote} 
( ``\textit{data quality}'' $\lor$ ``\textit{information quality}'') $\land$ \textit{tool}
\end{quote}

\begin{table}[ht]
\caption{Systematic Search}
\label{tab:systematicsearch}
\footnotesize
\begin{center}
\begin{tabular}{p{2.9cm}p{7.8cm}p{1.9cm}p{2.3cm}}
  \hline
  Source & Search Expression & Scope & Restrictions\\ 
  \hline
  \rowcolor{lightgray} 
  ACM Digital Library\tablefootnote{\url{http://dl.acm.org/advsearch.cfm} (June 2019)}  & acmdlTitle:(+(``data quality'' ``information quality'') +tool) OR recordAbstract:(+(``data quality'' ``information quality'') +tool) & Title, abstract & - \\
  GitHub\tablefootnote{\url{https://github.com} (June 2019)} & ``data quality'' OR ``information quality'' & Full text& - \\
  \rowcolor{lightgray} 
  Google Scholar\tablefootnote{\url{https://scholar.google.at} (June 2019)} & allintitle: (``data quality'' OR ``information quality'') AND tool & Title & Exclude citations and patents \\
  IEEE Xplore Digital Library\tablefootnote{\url{http://ieeexplore.ieee.org/search/advsearch.jsp} (June 2019)}  & (((``data quality'') OR ``information quality'') AND tool) & Title, abstract, indexing terms & - \\
  \rowcolor{lightgray} 
  Science Direct\tablefootnote{\url{http://www.sciencedirect.com} (June 2019)} & TITLE-ABSTR-KEY(``data quality'' OR ``information quality'') and TITLE-ABSTR-KEY(tool)[All Sources(Computer Science)].  & Title, abstract, keywords & Computer science only \\
  Springer Link\tablefootnote{\url{https://link.springer.com/advanced-search} (June 2019)} & tool NEAR (``data quality'' OR ``information quality'') & Full text & Computer science only \\
  \hline
\end{tabular}
\end{center}
\end{table}%

The search expression has then been applied to the list of digital libraries that is provided in Table \ref{tab:systematicsearch}. 
We also included the software development platform GitHub, because the purpose of this search is to select concrete tools. 
The original aim was to search all titles and abstracts from the computer science domain. 
However, since each digital library offers different search functionalities, we selected the closest search-engine-specific settings to reflect our original search aim. 
Table \ref{tab:systematicsearch} documents the deviations for each conducted search along with the ultimately utilized search expression, which is already formatted according to the guidelines of the respective search engine. 
For the GitHub search, we additionally omitted the search term \textit{tool}, because most GitHub results are obviously tools (except for empty repositories, code samples, or documentations). 

\begin{figure}
\centering
  \includegraphics[scale=0.5]{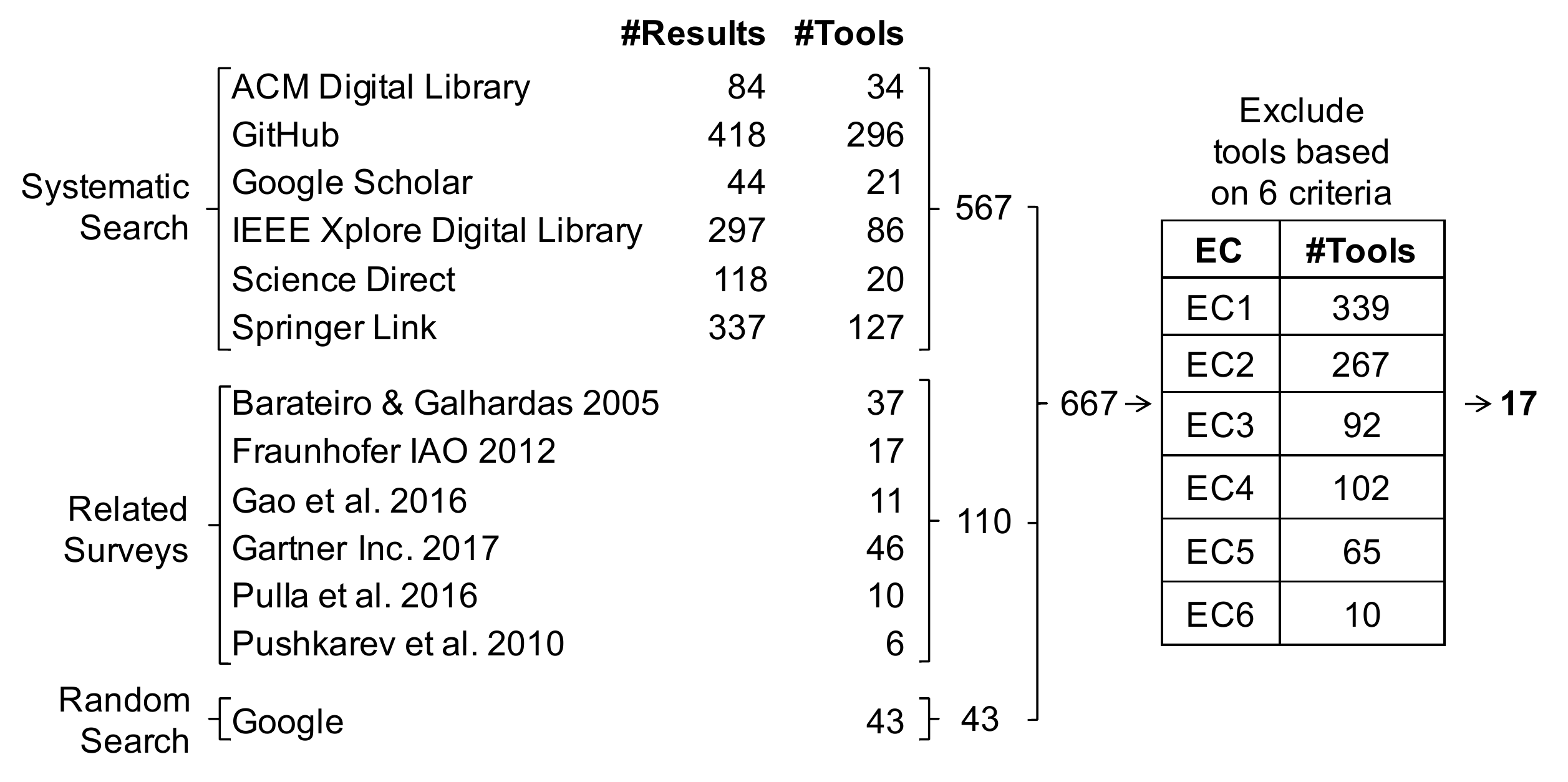}
  \caption{Systematic Search}
  \label{fig:SystematicSearch}
\end{figure}

For each search result, we assessed the title and abstract to determine whether a paper actually promotes a candidate DQ tool or not. 
In cases where title and abstract were not explicit enough, or they indicated the presentation of a tool (and therefore the paper could not be directly classified as not relevant), the content of the paper was investigated in more detail to record name and purpose of the tool in a first step. 
In the GitHub search, we excluded all tools that did not offer any kind of description immediately and used the others as candidates. 
Figure \ref{fig:SystematicSearch} illustrates the number of investigated research papers and the resulting tools. 
The next section describes the subsequent investigation of all candidate tools according to defined exclusion criteria (EC). 

\subsection{DQ Tool Selection}
In accordance with our general search strategy, we defined three inclusion criteria. Each tool that was selected as candidate tool had to satisfy at least one of the following three criteria.
\begin{enumerate}
	\item The tool was included in one of the previous surveys~\cite{GartnerDQTools2017, Barateiro_DQSurvey_2005, Pushkarev_DQSurvey_2010, Pulla_DQSurvey_2016, GaoEtAl_BigDataQA_2016, Fraunhofer2012}.
	\item The tool was identified in our systematic search. 
	\item The tool was identified in our random search. 
\end{enumerate}

Figure \ref{fig:SystematicSearch} shows the number of results (\#Results) as well as the number of tools (\#Tools) we found in the systematic search per source. In total, 1,298 results have been discovered through the systematic search of which 567 describe or refer to a DQ tool (this number contains research referring to the same tool, i.e., duplicates). In the related surveys we located 110 tools (incl. duplicates) and added 43 additional tools from the random Google search. In the next step, all these 720 tools have been listed in one spreadsheet file in order to remove duplicates. 
This resulted in a total of 667 identified distinct DQ tools. 
After establishing the list of candidate tools, we conducted a review to exclude all tools from the survey that met at least one of the following exclusion criteria. 

\begin{enumerate}[label=(EC\arabic*),leftmargin=1.4cm]
	\item The tool is domain-specific (e.g., for web data or a specific implementations only). 
	\item The tool is dedicated to specific data management tasks without explicitly offering DQ measurement.
	\begin{enumerate}
		\item The tool is dedicated to data cleansing.
		\item The tool is dedicated to data integration (including  on-the-fly DQ checks). 
		\item The tool is dedicated to other data management tasks (e.g., data visualization).
	\end{enumerate}
		\item The tool is not publicly available (e.g., the tool is only described in a research paper). 
	\item The tool is considered deprecated (i.e., the vendor does not exist any more or the tool was found on GitHub and the last commit was before January 1\textsuperscript{st}, 2016). 
	\item The tool was found on GitHub without any further information available. 
	\item The tool requires a fee and no free trial is offered upon request. 
\end{enumerate}

The table in Figure \ref{fig:SystematicSearch} shows how many tools were excluded per criterion (multiple selection was possible). Most of the tools were excluded because they are domain specific (EC1) and/or focus on specific data management tasks (EC2). The 267 tools excluded due to EC2 are divided between the three subcriteria as follows: 111 tools were excluded by EC2a, 46 tools were excluded by EC2b and 110 tools by EC2c.
17 DQ tools have been selected for deeper investigation, from which 13 could be evaluated since three were based on SAP, where no installation was available, and one (IBM InfoSphere Information Server) could not be installed successfully during the time of the project, despite great effort but with little support from IBM. 

\section{Theory on Data Quality and Evaluation Framework}
\label{sec:theory}
% General about DQ
Despite different existing interpretations, the term ``data quality'' is most frequently described as ``fitness for use''~\cite{Chrisman_1984, WangStrong_1996}, referring to the high subjectivity and context-dependency of this topic. Information quality is often used as synonym for data quality and even though both terms can be clearly distinguished, because ``data'' refers to plain facts and ``information'' describes the extension of those facts with context and semantics, they are often used interchangeably in the DQ literature~\cite{Wang_1998, ZhuLeeWang_2014}.
We use the term data quality because our focus is on processing objectively, automatically retrievable facts (i.e., intrinsic data characteristics). The term information serves as synonym for data in the systematic search to achieve higher coverage. 

% Explain DQ Program and rough overview
A data quality methodology can be divided in its core into the following activities~\cite{Batini_2009methodologies, Maydanchik_2007, English_1999}: (1) state reconstruction, (2) DQ measurement or assessment, (3) data cleansing or improvement, and (4) the establishment of continuous DQ monitoring. 
We want to point out that not all methodologies include all of those steps, for example, step (1) is omitted in~\cite{Maydanchik_2007} and step (4) is omitted in the DQ methodology survey in~\cite{Batini_2009methodologies}. Further, some methodologies include additional activities like monitoring of data integration interfaces (cf.~\cite{Maydanchik_2007}), which we do not consider because of their specialization. 
In the following paragraphs we describe the four core steps of a DQ methodology in detail to clarify the difference between DQ measurement, DQ monitoring, and data cleansing activities, where the latter ones are not included in the survey. Step (1), the state reconstruction, describes the collection of contextual information on the observed data as well as on the organization where a DQ project is carried out~\cite{Batini_2009methodologies}. Since the focus of this paper is on DQ tool functionalities, we restrict step (1) in the following to the data part (i.e., data profiling) and do not describe gathering of contextual information on the organization in detail. 

\paragraph{Data profiling.} 
Data profiling is described as the process of analyzing a dataset to collect data about data (i.e., metadata) using a broad range of techniques~\cite{Naumann_2014, Abedjan_2015, Abedjan2019}. Thus, it is an essential task prior to any DQ measurement or monitoring activity to get insight into a given dataset. 
Exemplary information that is gathered during data profiling are the number of distinct or missing (i.e., null) values in a column, data types of attributes, or occurring patterns and their frequency (e.g., formatting of telephone numbers)~\cite{Abedjan_2015}. We refer to~\cite{Abedjan_2015,Abedjan2019} for a detailed discussion on data profiling techniques and tasks. 
Acccording to Gartner~\cite{GartnerDQTools2017} and the findings of our survey, most general-purpose DQ tools offer data profiling capabilities to some extent. 

\paragraph{Data quality measurement.}
% DQ Measurement vs. Assessment
According to Sebastian-Coleman~\cite{Coleman_2012}, one of the biggest challenges for DQ practitioners is to answer the question on how data quality should be actually measured. Ge and Helfert~\cite{GeHelfert_2007} indicate that this is also true for the synonymously used term \textit{assessment} by stating that one of the major questions in DQ research is ``How to assess data quality?''. 
The term \textit{measure} describes ``to ascertain the size, amount, or degree of something by using an instrument or device marked in standard units or by comparing it with an object of known size''~\cite{NOAD}. Although the term \textit{assessment} is often used as synonym for measurement, especially in DQ literature there is a clear distinction between both terms. Assessment is the ``evaluation or estimation of the nature, ability, or quality of something'' and extends the concept of measurement by evaluating the measurement results and drawing a conclusion about the object of assessment~\cite{NOAD, Coleman_2012}. 
DQ assessment is also described as the detection and initial estimation of data quality as well as the impact analysis of occurring DQ problems~\cite{ApelEtAl_2015, English_1999}. 
In this survey, we use the term \textit{measurement} since the focus is on measurement capabilities of DQ tools, independently of the interpretation of the results by a user. 

Traditionally, DQ measurement is described by a set of DQ dimensions and assigned metrics. According to Lee et al.~\cite{Lee_2009Journey}, ``DQ assessment requires assessments along a number of dimensions''.
Despite the wide agreement on DQ dimensions in general and a lot of research over the last decades, there is still no consensus on a standardized list of dimensions and metrics for DQ measurement~\cite{Coleman_2012, CCDQ}. 
Thus, we observe existing DQ dimensions and metrics and justify their inclusion in our requirements catalog in Subsection~\ref{sec:dqdim}. 

\paragraph{Data cleansing.}
Data cleansing describes process of correcting erroneous data or data glitches. In practice, automatable cleansing tasks include customer data standardization, de-duplication, and matching. Other efforts to improve DQ are usually performed manually. 
While automated data cleansing methods are very valuable for large amounts of data, they pose risks to insert new errors that are rarely well understood~\cite{Maydanchik_2007}. 
%% False positives??
We intentionally did not observe data cleansing functionalities in this survey, since the focus is on the detection of DQ problems. However, data cleansing algorithms are usually based on DQ measurement, since it is initially necessary to detect DQ problems to increase the quality of a given dataset.

\paragraph{Data quality monitoring.}
The term ``DQ monitoring'' is mainly used implicitly in literature without an established definition and common understanding. This leads to different interpretations when the term is mentioned in scientific publications or by companies promoting and describing their DQ tool. There is a difference between ``data monitoring'', which describes continuous checking of rules, and ``DQ monitoring'', which is ongoing measurement of DQ~\cite{EhrlingerWoess_2017}. The aim of this survey is to observe not only the functionalities of current DQ tools in terms of data profiling and measurement, but also in terms of true DQ monitoring. Pushkarev et al.~\cite{Pushkarev_DQSurvey_2010} and a follow-up study~\cite{Pulla_DQSurvey_2016} point out that none of the tools observed had any monitoring functionality. We however want to include this criterion in our requirements catalog since there is evidence on several DQ tool websites that they do offer monitoring functionalities, but have not been observed by~\cite{Pushkarev_DQSurvey_2010} and~\cite{Pulla_DQSurvey_2016}. 

\subsection{Data Quality Dimensions and Metrics}
\label{sec:dqdim}
Data quality is often described as concept with multiple dimensions, so that every DQ dimension refers to a specific aspect of the quality of data~\cite{Ehrlinger_Minimality}. 
Over the years, a wide variety of dimensions and dimension classifications have been proposed~\cite{BallouPazer_1985, Pipino_DQA_2002, GeHelfert_2007, WangStrong_1996, BatiniScannapieco_2016, WandWang_1996}. An overview of possible dimensions and classifications is provided in~\cite{LaranjeiroEtAl_2015, ScannapiecoCatarci_2002}. 
Despite intensive research and an ongoing discussion on DQ dimensions, there is still no consensus on which dimensions are the essence for DQ measurement~\cite{Coleman_2012}. 
Our evaluation framework covers the four most frequently used dimensions, which are according to~\cite{ScannapiecoCatarci_2002, Hildebrand_2015, WandWang_1996}: accuracy, completeness, consistency, and timeliness. 

Piro~\cite{Piro_2014} distinguishes between ``hard dimensions'' (including accuracy, completeness, and timeliness, amongst others), which can be measured objectively using check routines, and ``soft dimensions'', which can only be assessed using subjective evaluation. However, also objective check routines require a preceding subjective and domain-specific definition of the data objects to be measured, in order to consequently follow the ``fitness for use'' approach~\cite{Piro_2014}. 

In conjunction with the discussion of DQ dimensions, it is often mentioned that the definition of specific DQ metrics is required to apply those dimensions in practice. A metric is a function that maps a quality dimension to a numerical value, which allows an interpretation of a dimension's fulfillment~\cite{IEEEStandard_QualityMetric}. 
Such a DQ metric can be measured on different aggregation levels: on value-level, column or attribute-level, tuple or record-level, table or relation-level, as well as database (DB)-level~\cite{Hildebrand_2015}. The aggregation could, for example, be performed with the weighted arithmetic mean of the calculated metric results from the previous level (e.g., results of the record-level to calculate the table-level metric)~\cite{Hinrichs_2002}. 
Heinrich et al.~\cite{Heinrich_2018} proposed five requirements for DQ metrics to ensure reliable decision-making: ``the existence
of minimum and maximum metric values (R1), the interval scaling of the metric values (R2), the quality of the configuration parameters and the determination of the metric values (R3), the sound aggregation of the metric values (R4), and the economic efficiency of the metric (R5)''~\cite{Heinrich_2018}. 
However, other researchers claim ``that a more general approach is required''~\cite{Bronselaer_2018} to assess the usefulness and validity of a DQ metric. 
In the following, we describe four prominent DQ dimensions along with common metrics for their calculation. The list of metrics is not exhaustive, but should give an impression about the research conducted in this area since we observe the existence of such or similar metrics in our DQ tool evaluation. 

\subsubsection{Accuracy} 
Although accuracy is sometimes described as the most important data quality dimension, a number of different definitions exist~\cite{WandWang_1996, Haegemans_2016Accuracy}. 
In DQ literature, accuracy can be described as the closeness between an information system and the part of the real-world it is supposed to model~\cite{BatiniScannapieco_2016}. From the natural sciences perspective, accuracy is usually defined as the ``magnitude of an error''~\cite{Haegemans_2016Accuracy}. 
We refer to Haegemans et al.~\cite{Haegemans_2016Accuracy} for a detailed discussion on the definitions of accuracy and a comprehensive list of published metrics related to accuracy. 
Here, we provide three exemplary metrics: 
\begin{equation}\label{eq:accuracy1}
\textit{Free-of-error rating} = 1 - \frac{\textit{Number of data units in error}}{\textit{Total number of data units}}~\cite{Lee_2009Journey},
\end{equation}

which was published by Lee et al.~\cite{Lee_2009Journey}, and a modification of Equation (\ref{eq:accuracy1}) that takes into account the randomness of the occurrence of an error $ROE$ and the probability distribution of the occurrence of an error $PDOE$~\cite{Fisher_Accuracy_2009}: 
\begin{equation}\label{eq:accuracy2}
\textit{accuracy} = \left( \frac{\textit{NrOfCorrectValues}}{\textit{TotalNrOfValues}}, \textit{ROE}, \textit{PDOE} \right)~\cite{Fisher_Accuracy_2009}.
\end{equation}

Hinrichs~\cite{Hinrichs_2002} proposed the accuracy metric in Equation~\ref{eq:accuracy3}, which can be aggregated on different levels. On attribute-value-level, the metric $Q_{Gen}$ for accuracy ($Gen$ is German for ``Genauigkeit'', english for accuracy) is defined by the ratio between a value's arity and its optimal arity for numeric values. For a numeric attribute $A$, $s_{opt}(A)$ is the optimal number of digits and decimals for $A$, $w$ is a value of $A$ and $s(w)$ is the actual number of digits and decimals for $w$ in attribute $A$. Since $s_{opt}(A)$ is not necessarily maximal, the metric needs to be normalized by $]0,1]$~\cite{Hinrichs_2002}.
\begin{equation}\label{eq:accuracy3}
Q_{Gen}(w,A) =  min \left( \frac{s(w)}{s_{opt}(A)},1 \right)~\cite{Hinrichs_2002}.
\end{equation}

For non-numeric attributes, Hinrichs~\cite{Hinrichs_2002} suggests to assign $w$ to plane $i$ within a classification $K$ with $n$ planes $(K_1,...,K_n)$ and to replace $s(w)$ with $i$ and to select $s_{opt}(A)$ from $K$ with $s_{opt}(A) \leq n$.
For a tuple $t$, accuracy $Q_{Gen}$ is measured according to: 
\begin{equation}\label{eq:accuracy4}
Q_{Gen}(t) =  \frac{\sum_{j=1}^{n} Q_{Gen}(t.A_{j}, A_{j})g_{j}}{\sum_{j=1}^{n}g_{j}}~\cite{Hinrichs_2002},
\end{equation}

where $t.A_{1},...,t.A_{n}$ are the attribute values for attributes $A_{1},...,A_{n}$ that specify  the observed tuple $t$. Factor $g_{j}$ is the relative importance of $A_{j}$ with respect to the total tuple and is an expert-defined weight~\cite{Hinrichs_2002}. The accuracy on table-level is then calculated as the arithmetic mean of the tuple accuracy measurements, and the accuracy on DB-level is the arithmetic mean of the table-level accuracy measurements. For a more detailed discussion on the metric, we refer to~\cite{Hinrichs_2002}. 

\subsubsection{Completeness} 
Completeness is very generally described as the ``breadth, depth, and scope of information contained in the data''~\cite{WangStrong_1996, BatiniScannapieco_2016} and covers the condition for data to exist to be complete. 
A common metric for completeness is provided by Lee et al.~\cite{Lee_2009Journey}:
\begin{equation}\label{eq:comp2}
\textit{completeness} =  1-\frac{\textit{number of incomplete elements}}{\textit{total number of elements}}~\cite{Lee_2009Journey}. 
\end{equation}

In other metrics (cf.~\cite{BatiniScannapieco_2016}), the number of complete elements is interpreted as the number of elements in a true reference set and the $-1$ is omitted. It should be pointed out that the number of missing values can be calculated in different ways, either by taking into account only true missing values (i.e. \texttt{null}), or also default values or a textual entry stating ``NaN'' (i.e., not a number). 
Hinrichs~\cite{Hinrichs_2002} assigns $0.0$ to a field value that is \texttt{null} or equivalent and $1.0$ else, and calculates completeness analogously to the accuracy metric on different aggregation levels (e.g., table-level or DB-level completeness) with the weighted arithmetic mean. 

\subsubsection{Timeliness} Timeliness describes ``how current the data are for the task at hand''~\cite{BatiniScannapieco_2016} and is closely connected to the notions of \textit{currency} (update frequency of data) and \textit{volatility} (how fast data becomes irrelevant)~\cite{BatiniScannapieco_2016}. A different definition states that ``timeliness can be interpreted as the probability that an attribute value is still up-to-date''~\cite{HeinrichKaiserKlier_2007}. 
A list of different metrics to calculate timeliness is provided in~\cite{Heinrich_2009Timeliness}, where the authors suggest calculating timeliness based on the definition in~\cite{HeinrichKaiserKlier_2007} according to:
\begin{equation}\label{eq:timeliness}
Q_{Time.}^{\omega} \left( t \right) := \text{exp} \left( -\textit{decline}\left( A \right) \cdot t \right)~\cite{HeinrichKaiserKlier_2007},
\end{equation}

where $\omega$ is the considered attribute value and $\textit{decline}\left( A \right)$ is the decline rate, which specifies the average number of attributes that become outdated within the time period $t$~\cite{HeinrichKaiserKlier_2007}. 

\subsubsection{Consistency} There are also different definitions for the consistency dimension. According to Batini and Scannapieco~\cite{BatiniScannapieco_2016}, ``consistency captures the violation of semantic rules defined over data items, where items can be tuples of relational tables or records in a file''~\cite{BatiniScannapieco_2016}. An example for such rules are integrity constraints from the relational theory. Hinrichs~\cite{Hinrichs_2002} assumes for his proposed consistency metric that domain knowledge is encoded into rules and excludes contradictions within the rules and fuzzy or probabilistic assumptions. Consequently the consistency $Q_{Kon}$ ($Kon$ is German for ``Konsistenz'', English for consistency) of an attribute value $w$ is defined by~\cite{Hinrichs_2002} as 
\begin{equation}\label{eq:consistency1}
Q_{Kon}(w) =  \frac{1}{\sum_{j=1}^{n}r_{j}(w)g_j+1}~\cite{Hinrichs_2002},
\end{equation}

where $g_j$ is the degree of severity of $r_j(w)$, and $r_j(w)$ is the violation of consistency rule $r_j$ (within a set of $n$ consistency rules), applied to the attribute value $w$, and defined as 
\begin{equation}\label{eq:consistency2}
r_j(w)
\begin{cases}
    0  & \text{if } w \text{ satisfies } r_j\\
    1  & \text{otherwise.}~\cite{Hinrichs_2002}
\end{cases}
\end{equation}

Consistency rules cannot only be defined on attribute-value-level, but also on tuple-level. The calculation of the consistency on table- or database-level is an alignment to the accuracy and completeness metric calculated as the arithmetic mean of the tuple-level consistency~\cite{Hinrichs_2002}. 

Sebastian-Coleman~\cite{Coleman_2012} suggests measuring consistency over time by comparing the ``record count distribution of values (column profile) to past instances of data populating the same field''~\cite{Coleman_2012}. 

This list of metrics for the DQ dimensions accuracy, completeness, timeliness, and consistency, is by no means exhaustive, but a comprehensive discussion would be out of scope for this paper. We conclude that literature offers a number of specifically formulated metrics to measure DQ dimensions and this survey observes their existence in state-of-the-art DQ tools. 

\subsection{Requirements for Data Quality Tools}
Although existing literature offers different requirements and requirement classifications a DQ must fulfill, those were not sufficient to answer our research questions due to two major reasons: (1) they were too general for our evaluation, and (2) they had a different focus in terms of functionality. 

An example for a DQ tool requirement description with too little details is~\cite{GartnerDQTools2017}, where Gartner observes the following DQ capabilities: connectivity, data profiling, measurement and visualization, monitoring, parsing, standardization and cleaning, matching, linking and merging, multidomain support, address validation/geocoding, data curation and enrichment, issue resolution and workflow, metadata management, DevOps environment, deployment environment, architecture and integration, and usability. 
Similarly, Loshin~\cite{Loshin_2010} defines the following eight requirements a DQ tool must offer: ``data profiling, parsing, standardization, identity resolution, record linkage and merging, data cleansing, data enhancement, and data inspection and monitoring''~\cite{Loshin_2010}. 
In contrast to such evaluations, we focused specifically on data profiling, DQ measurement, and DQ monitoring functionality. While general features like connectivity and usability of the tools are not necessary to answer our research questions, we added a short textual description to each tool we observed. 

The differentiation of our DQ tool survey to existing ones (and consequently their requirements) is already explained in detail in Section~\ref{sec:relSurveys}. In summary, the proposed requirements were of too less detail or with a different functional focus. 
In addition to existing surveys, Goasdou\'{e} et al.~\cite{Goasdoue_2007} explicitly proposed an evaluation framework for DQ tools without publishing the results of their evaluation, which is why they have not been included in our list of related surveys. 
However, we did not use their criteria as proposed, since the requirements were adapted to the context of the company, they performed the DQ tool evaluation for: \'{E}lectricit\'{e} de France (EDF), a French electric utility company, and more precisely to their CRM (customer relationship management) environments. 
The main differences to our requirements catalog are a more detailed evaluation of address normalization, duplicate detection, and reporting capabilities, but less details in data profiling and no coverage of DQ monitoring functionality.  

\begin{table}[ht]
\caption{DQ Tool Requirements Catalog}
\label{tab:reqCatalog}
\footnotesize
\begin{center}
\begin{tabular}{p{1.8cm}p{3cm}l}
  \toprule
  Category & Sub-category & Requirement \\ 
  \midrule
  \multirow{25}{*}{\parbox{1.8cm}{Data Profiling  \cite{Abedjan_2015, Abedjan2019}}} & SC -- Cardinalities & (1) Number of rows \\
  & & (2) Number of null values \\
  & & (3) Percentage of null values \\
  & & (4) Number of distinct values; sometimes called ``cardinality'' \\
  & & (5) Number of distinct values divided by the number of rows \\
  & \multirow{2}{*}{\parbox{3cm}{SC - Value distributions}} & (6) Frequency histograms (equi-width, equi-depth, etc.) \\
  & & (7) Minimum and maximum values in a numeric column \\ 
  & & (8) Constancy: Frequency of most frequent value divided by number of rows \\ 
  & & (9) Quartiles: 3 points that divide the (numeric) values into 4 equal groups \\
  & & (10) Distribution of first digit in numeric values; to check Benford's law\\
  & \multirow{2}{*}{\parbox{3cm}{SC - Patterns, data types, and domains}} & (11) Basic type (e.g., numeric, alphanumeric, date, time) \\
  & & (12) DBMS-specific data type (e.g., varchar, timestamp) \\
  & & (13) Measurement of value length (minimum, maximum, average, and median) \\
  & & (14) Maximum number of digits in numeric values \\
  & & (15) Maximum number of decimals in numeric values \\
  & & (16) Histogram of value patterns (Aa9...) \\
  & & (17) Generic semantic data type (e.g., code, date/time, quantity, identifier) \\
  & & (18) Semantic domain (e.g., credit card, first name, city) \\
  & Dependencies & (19) Unique column combinations (key discovery) \\
  & & (20) Relaxed unqiue column combinations \\
  & & (21) Inclusion dependencies (foreign key discovery) \\
  & & (22) Relaxed inclusion dependencies \\
  & & (23) Functional dependencies \\
  & & (24) Conditional functional dependencies \\
  & \multirow{2}{*}{\parbox{3cm}{Advanced MC profiling}} & (25) Correlation analysis \\
  & & (26) Association rule mining \\
  & & (27) Cluster analysis \\
  & & (28) Outlier detection \\
  & & (29) Exact duplicate tuple detection \\
  & & (30) Relaxed duplicate tuple detection \\
  \midrule
  \multirow{8}{*}{\parbox{1.8cm}{Data Quality Measurement}} & DQ Dimensions & (31) Metric to measure accuracy\\
  & & (32) Metric to measure completeness \\
  & & (33) Metric to measure consistency \\
  & & (34) Metric to measure timeliness \\
  & & (35) Metrics to measure other DQ dimensions \\
  & Rule-based checks & (36) Creation of business rules \\
  & & (37) Availability of general-applicable integrity rules \\
  & & (38) Verification of data against business rules \\
  \midrule
  \multirow{5}{*}{Continuous Data Quality Monitoring \cite{EhrlingerWoess_2017}} & & (39) Scheduling a DQ metric or data profiling task in user-defined periods  \\
  & & (40) Storage of DQ measurements and data profiling results  \\
  & & (41) Retrieval of DQ measurements or data profiling results \\
  & & (42) Comparison between several DQ measurements or data profiling results  \\
  & & (43) Visualization of DQ measurements / data profiling results over time  \\
  \bottomrule
\end{tabular}
\end{center}
\end{table}%

\subsection{Evaluation Requirements Catalog}
\label{sec:req_catalog}
Based on the theory introduced in the preceding paragraphs, we developed a DQ tool requirements catalog (see Table~\ref{tab:reqCatalog}) for our evaluation, which consists of three main categories: data profiling (DP), data quality measurement (DQM), and continuous data quality monitoring (CDQM). 
The requirements for data profiling are based on the classification of DP tasks by Abedjan et al., which has been originally published in~\cite{Abedjan_2015}, and recently updated in~\cite{Abedjan2019}. Since we started our survey prior to the classification update, our requirement catalog constitutes a tradeoff between the two versions. Since both versions contain the two sub-categories ``single columns (SC) profiling'' and ``dependency detection'', we adhere here to the newer version in~\cite{Abedjan2019}. In the SC sub-category, we split the null values task (i.e., number or percentage of null values) in two different requirements: (DP-2) number of null values and (DP-3) percentage of null values, to separate the results. 
The newer version~\cite{Abedjan2019} contains an additional sub-category ``metadata for non-relational data,'' which is not included in our survey, because the evaluation for some tools with a fixed-period trial version was already completed at the time of the update. However, the original version~\cite{Abedjan_2015} included a category ``multi-column (MC) profiling'', which has been removed in~\cite{Abedjan2019}. We renamed this category to ``advanced MC profiling'' and added it along with two additional requirements (exact and relaxed duplicate tuple detection) to the end of the DP category. One reason for the exclusion of the MC sub-category from the data profiling task taxonomy in~\cite{Abedjan2019} might be the strong overlap of these tasks with the field of data mining. Abedjan et al.~\cite{Abedjan2019} point out that there exists no clearly defined and widely accepted distinction between the two research fields. Thus, although a separate category for those requirements could be argued, we decided to include it in the data profiling category, because data mining is not in the focus of our survey. 

The category for DQ measurement contains requirements to provide metrics for specific DQ dimensions and business rule management capabilities. While we listed metrics for the DQ dimensions accuracy, completeness, consistency, and timeliness as described in Section~\ref{sec:dqdim} explicitly, we investigate the existence of additional metrics during our evaluation by means of requirement (DQM-34). 
There is no agreed-on metric for the integrity dimension as defined in~\cite{Coleman_2012}, but it can be checked with a set of rules. Thus, we distinguish in terms of rule-based checks between the (DQM-35) creation of domain-specific business rules and the (DQM-36) availability of general integrity rules, for example a birth date cannot be in the future or a temperature value can never reach -270 \degree C. It should also be possible to verify those rules (DQM-37).

The requirements for CDQM are based on the findings published in~\cite{EhrlingerWoess_2017} and summarize key tasks to ensure continuous DQ measuring over time. The continuous measurement, storage, and usage of the collected metadata should be possible for both data profiling results and DQ measurements. 

In addition to the requirements in Table~\ref{tab:reqCatalog}, we performed defined test cases for each requirement and compared the respective results from different tools. The basis for this evaluation is a modernized version of the well-known Northwind DB  by dofactory (cf.~Appendix A for the DB and test cases). 
\section{Data Quality Tool Evaluation}
\label{sec:eval}
In this section, we first describe the DQ tools, which we selected for the evaluation, and second, we investigate the selected tools with respect to our evaluation framework and discuss the requirements. 

\subsection{Selected Data Quality Tools}
\label{sec:tool_descr}
In total, we selected 17 DQ tools for detailed evaluation. Three of them were based on SAP (SAP Information Steward, DQ solution by ISO Professional Services, and dspCompose by BackOffice Associates GmbH) and since we had no access to a SAP installation, we did not include those tools in our survey, but described them textually. 
In the next subsections, we describe (in alphabetical order) each of the selected tool in more detail by specifically answering the following questions: 
\begin{itemize}
    \item Where and how was the tool discovered in our search? Which other surveys investigated it?
    \item Who is the vendor or creator of the tool? Is it open source or not? Does it require a fee?
    \item Which exact version did we evaluate, and how was it provided? (e.g., 30-days trial)
    \item What is the overall focus of the tool in terms of DQ functionality?
    \item How did we perceive the usability and customer support? 
\end{itemize}

\subsubsection{Aggregate Profiler}
Aggregate Profiler (AP) is a freely available and open-source DQ tool, which is dedicated to data profiling. The tool was discovered twice in our systematic search: once because it was mentioned by Dai et al.~\cite{Dai2016} in the Springer search results, and once in the Google search results, since it is also published on Sourceforge as ``Open Source Data Quality and Profiling''\footnote{\url{https://sourceforge.net/projects/dataquality} (June 2019)}, developed by \textit{arrah} and \textit{arunwizz}. We evaluated Aggregate Profiler version 6.2.4.
In addition to its data profiling capabilities, like statistical analysis and pattern matching, Aggregate Profiler can also be used for data preparation and cleansing activities, like address correction or duplicate removal. Moreover, business rules can be defined and scheduled in user-defined periods. 
We perceived the user interface (UI) as average, the design is inferior compared to other tools and the handling of some functions was not intuitive. The overall experience was perceived as average.

\subsubsection{Apache Griffin}
The free and open-source tool Apache Griffin\footnote{\url{https://griffin.incubator.apache.org} (June 2019)} (AG) was included in our GitHub search results and differs significantly from the other tools in this survey, because it does not offer any data profiling functionality and is not a comprehensive DQ solution. However, since part of the evaluation is to observe the extent to which current tools support CDQM, we included Apache Griffin, since it is dedicated to continuously measure the quality of Big Data, both batch-based and streaming data. We installed Apache Griffin 0.2.0, which is still in the incubator status of Apache, on Ubuntu 18.04. The tool requires the following dependencies, from which some are (at the time of the installation) still in incubating status as well: JDK (1.8+), MySQL DB, npm, Hadoop (2.6.0+), Spark (2.2.1+), Hive (2.2.0), Livy, and ElasticSearch. Due to these dependencies, the installation was very cumbersome in contrast to other tools. In our case, two experienced computer scientists needed over a week to complete the full installation. Once installed, the UI is intuitive and supports the domain-specific definition of accuracy metrics as well as the scheduling and monitoring of those metrics. Other DQ metrics, like completeness, are planned to be integrated in future versions. 

\subsubsection{Ataccama ONE}
The company Ataccama with its headquarters in Canada offers several DQ products, which we found through different sources in our search: Data Quality Center and Master Data Center have been previously investigated by Fraunhofer IAO~\cite{Fraunhofer2012}; DQ Analyzer has been included in the surveys~\cite{Pushkarev_DQSurvey_2010} and~\cite{Pulla_DQSurvey_2016}. Gartner additionally mentioned the DQ Issue Tracker and the DQ Dashboard in 2016~\cite{GartnerDQTools2016}. However, since 2017, Ataccama consolidated their separate DQ solutions into ``Ataccama ONE'' (A-ONE). 
While the licence of the full DQ solutions is subject to costs, the data profiling module of Ataccama ONE can be accessed freely. Unfortunately, Ataccama customer support did not provide us with a trial licence of the complete ONE solution. Thus, we were only able to investigate the free ``Ataccama ONE profiler''\footnote{\url{https://one.ataccama.com} (June 2019)}, where the focus is on data profiling and which does not provide monitoring functionality. We performed the evaluation of the online-available tool during October 2018. According to Gartner~\cite{GartnerDQTools2017} and Ataccama customer support, the full solution would provide a much richer scope of functions, including DQ monitoring, but we were not able to investigate it. 
The data profiling module was very intuitive and easy to use, also for business users. In terms of customer support (from Prague), we experienced very long response times on our contact attempts for a licence request. Additionally, we were promised to receive a training as prerequisite to test the full Ataccama ONE solution, which was never redeemed due to the workload on the side of Ataccama.  

\subsubsection{DataCleaner by Human Inference}
The DQ products ``DataCleaner'' (DC) and ``DataHub'' were originally developed by Human Inference, which was incorporated into Neopost in 2012, later into Quadient, and since 2019 into the EDM Media Group, where it is again promoted with its original name ``Human Inference''. Our customer contact declared that the professional version of DataCleaner (in contrast to the community edition that is freely avaiable online) offers the same DQ measurement functionalities as DataHub, but differs only with respect to the convenient usage, the UI, and the data integration features. Thus, we evaluated a trial version of DataCleaner Enterprise Edition version 6.3.0, which aims at people with technical background. In addition, we were able to observe the functionalities of DataHub in an interactive web session. The focus of Human Inference is customer data, which is reflected in specialized algorithms for duplicate detection, address matching, and data cleansing.
Under the vendor Quadient, DataCleaner was mentioned by Gartner in 2017~\cite{GartnerDQTools2017}, but excluded from the follow-up survey in 2019~\cite{GartnerDQTools2019}, due to strategic changes.
Although DataCleaner is built for technical users, we perceived the UI as very intuitive. DataHub (with its vision of a single customer view) offers in addition to the administrator's view a restricted data steward view, which is specifically dedicated to business users, for example, to resolve ambiguous duplicates.
We also want to highlight (in conformance with~\cite{GartnerDQTools2017}) the very helpful and friendly customer support that provided us with the trial licence and more insight in DataHub. 

\subsubsection{Datamartist by nModal Solutions Inc.}
The commercial tool Datamartist (DM) by nModal Solutions Inc. was already mentioned by Pulla et al.~\cite{Pulla_DQSurvey_2016} and additionally discovered in our Google search. We investigated the free-to-download 30-days trial of Datamartist version 1.7.9, which offers the complete Pro edition features. Installation requirements are the operating system Microsoft Windows and the .NET framework 2.0. Datamartist is dedicated to data profiling and data transformation. We perceived the UI of Datamartist as slightly inferior compared to other commercial tools and for some tasks (e.g., export data profiling results) the command line was required. Since the trial could be downloaded from the website directly, we did not consult any customer support.

\subsubsection{Experian Pandora}
The company Experian with its headquarters in Ireland offers two commercial DQ solutions: Cleanse and Pandora (EP). During the conduct of our survey, they introduced the new product Aperture Data Studio, which is going to replace Pandora in the future. In our systematic search, we found the Experian DQ tools only through the Gartner studies~\cite{GartnerDQTools2017, GartnerDQTools2016} and did not find any other mention in related work. While Cleanse is dedicated to one-time-data-cleansing, we investigated the more comprehensive tool Pandora version 5.9.0 in a 30-days free trial. In accordance with the findings by Gartner~\cite{GartnerDQTools2017}, we perceived the tool as easy to install and use and want to highlight the comprehensive data profiling capabilities in general, and the cross-table profiling capabilities in particular. In addition, Pandora provides a rich ability to extend the existing feature palette with customized functions. In summary, Pandora achieved one of the best overall assessments in our survey. We perceived the UI as good, though more dedicated to technical users, and had very good experience with the technical customer support who supported us in a timely and target-oriented fashion. 

\subsubsection{Informatica Data Quality}
% \footnote{https://www.informatica.com (June 2019)}
Informatica Data Quality (IDQ) is one module of the commercial data management solution by Informatica, which is according to Gartner~\cite{GartnerDQTools2017, GartnerDQTools2016, GartnerDQTools2019}, leader in the Magic Quadrant of Data Quality Tools for several years. We were provided with two 30-days trial licences. The trial included the Informatica Developer (the desktop installation for developers), Informatica Analyst (the web-based platform for business users), and Informatica Administrator (for task scheduling), where all three user interfaces access the same server-side backend of Informatica DQ version 10.2.0. 
In our systematic search, we found five different tools offered by the company Informatica, from which four had been excluded from the evaluation. For example, the ``Master Data Management'' solution was excluded due to the focus on master data management. Informatica Data Quality was found through the Springer Link search, and because it was previously investigated by Gartner~\cite{GartnerDQTools2017,GartnerDQTools2016} and Gao et al.~\cite{GaoEtAl_BigDataQA_2016}. 
Informatica has its origin in the field of data integration and in addition to the features we evaluated, they offer data cleansing and matching functionalities. In terms of DQ measurement, they offer most probably the closest implementation to the DQ dimension and metric view promoted in the research community. 
We perceived the UI of Informatica Analyst as easy to use, also for business users, but with less comprehensive functionality than the Informatica Developer, which is more powerful and dedicated to trained and technical users. 
In accordance to the findings by Gartner customers~\cite{GartnerDQTools2017}, we can confirm the very helpful sales support, which was one of the best we experienced. During the evaluation, we had regular web conferences to ask questions and review the results, and short intermediate requests were answered timely. 

\subsubsection{IBM InfoSphere Information Server for Data Quality}
The product ``Infosphere Information Server for Data Quality'' (IBM ISDQ) by IBM was found through the studies by Gartner~\cite{GartnerDQTools2017,GartnerDQTools2019} and Fraunhofer IAO~\cite{Fraunhofer2012}. Other product (or product components) from IBM have also been previously mentioned in the following research papers: IBM Informix (previously called ``DataBlade'') by Barateiro and Galhardas~\cite{Barateiro_DQSurvey_2005}, IBM InfoSphere Information Analyzer by Abedjan et al.~\cite{Abedjan_2015}, IBM QuerySurge by Gao et al.~\cite{GaoEtAl_BigDataQA_2016}, IBM Data Integrator by Chen et al.~\cite{Chen2012}, and IBM InfoSphere MDM Server by Pawluk~\cite{Pawluk2010}. 
For our survey, the IBM partner solvistas GmbH, located in Austria, provided us with the installation files of IBM InfoSphere Information Server for Data Quality version 11.7 for a three-month trial. Unfortunately, we were not able to evaluate the tool due to an early error in the installation process stating that a required file was not found. Despite intensive research of the documentation\footnote{\url{https://www.ibm.com/support/knowledgecenter/en/SSZJPZ_11.7.0/com.ibm.swg.im.iis.productization.iisinfsv.install.doc/topics/cont_iisinfsrv_install.html} (June 2019)}, it was not possible to resolve the issue within the timeframe of the project, since no support by IBM nor any specific installation instruction for the received files was provided. We also contacted Fraunhofer IAO, who included IBM ISDQ in their survey~\cite{Fraunhofer2012}. However, they did not install the tool, but based their statements on contact with the IBM support and the documentation. Also solvistas GmbH claimed that, so far, they never installed the IBM DQ product line.
This experience aligns with the statement by Gartner that reference customer rate the technical support and documentation of IBM below the average~\cite{GartnerDQTools2019}.
% ``The image properties file was not found. The file is part of the entitlement package which can be obtained from your installation media or download site. The package name has the form Bundle*.zip. Before installation, extract contents of this package into the /is-suite directory''. 

\subsubsection{InfoZoom by humanIT Software GmbH}
% \footnote{http://www.humanit.de (June 2019)}
InfoZoom is a commercial DQ tool by the German vendor humanIT Software GmbH and is dedicated to data profiling using in-memory analytics. It was previously mentioned and investigated by Fraunhofer IAO in~\cite{Fraunhofer2012}, which was also the only source where we found the tool. 
We investigated InfoZoom Desktop Professional 2018 release 9.20 along with the IZDQ (InfoZoom Data Quality) extension version 2018.03 in a 6-month license (10.12.2018-30.06.2019) granted to us from the customer support. While InfoZoom Desktop is dedicated to data profiling and data investigation, the IZDQ extension allows a user to define rules and jobs for comprehensive DQ management. Generally, InfoZoom aims at observing and understanding the data but does not support any cleansing activities, which aligns well with the observations performed in this survey. We perceived the UI of InfoZoom Desktop as easy to use, also for business users, whereas the more advanced extension IZDQ requires technical knowledge like the ability to write SQL statements, or at least, intensive training to be used by non-technical users. 
The customer support was very friendly and helpful and provided us in a timely manner with a relatively long trial licences in comparison to other commercial DQ tools. 

\subsubsection{MobyDQ}
MobyDQ\footnote{\url{https://github.com/mobydq/mobydq} (June 2019)}, which was previously termed ``Data Quality Framework'', by Alexis Rolland is a free and open-source DQ solution that aims to automate DQ checks during data processing, storing DQ measurements and metric results, and triggering alerts in case of anomaly. 
The tool was inspired by an internal DQ project at Ubisoft Entertainment, which differs to the open-source version with respect to software dependency and mature but context-dependent configuration. 
We found MobyDQ through our GitHub search and evaluated the version downloaded on May 21, 2019. Similar to the commercial tools we observed, the framework can be used to access different data sources. 
In contrast to Apache Griffin, MobyDQ could be installed quickly and straightforward, based on the detailed documentation provided on GitHub.
MobyDQ does not provide any data profiling functionality, because its focus is on the creation, application, and automation of DQ checks. The creator Alexis Rolland was very helpful in demonstrating the productive installation at Ubisoft Entertainment to us, which clearly demonstrates the potential of the tool when applied in practice. 
% No version number provided

\subsubsection{OpenRefine and MetricDoc}
% \footnote{http://openrefine.org (June 2019)}
OpenRefine (formerly Google Refine, abbrev. OR) is a free and open-source DQ tool dedicated to data cleansing and data transformation and was discovered through~\cite{Kusumasari_2016} in the IEEE search results, and~\cite{Tsiflidou_2013} in the Springer Link search results as well as on GitHub\footnote{\url{https://github.com/OpenRefine/OpenRefine} (June 2019)}. While the original functionality of the tools does not primarily align with the focus of our survey, its extension MetricDoc specifically aims at assessing DQ with ``customizable, reusable quality metrics in combination with immediate visual feedback''~\cite{Bors_2018}.
Apart from the mention in~\cite{Kusumasari_2016, Tsiflidou_2013}, OpenRefine was not evaluated in one of the previous surveys on DQ tools, although it is open source. 
We installed the tool from GitHub and evaluated OpenRefine version 3.0 with the MetricDoc extension (where no version was provided), downloaded on February 14th, 2019. 
We perceived the usability of OpenRefine as average and especially in the MetricDoc extension, the usability of several functions reflected its state as very current research project. 

\subsubsection{Oracle Enterprise Data Quality}
The commercial tool Oracle Enterprise Data Quality (EDQ) was previously mentioned by Gartner~\cite{GartnerDQTools2017} and also found in~\cite{Abedjan_2015}, which was included in the Springer Link search results. 
We investigated the freely available pre-built Virtual Machine of version 12.2.1 available at the Oracle website\footnote{\url{http://www.oracle.com/technetwork/middleware/oedq/downloads/edq-vm-download-2424092.html} (June 2019)}. 
In addition to classical data profiling capabilities, EDQ offers data cleansing (parsing, standardization, match and merge, address verification), as well as DQ monitoring to some extent. 
The GUI was perceived as average with the major drawback being the inflexible data source connection to DBs and files. 
In comparison to other DQ tools, where a connection can be directly accessed and reused, Oracle EDQ requires a ``snapshot'' of the actual data connection to be created prior to any profiling or DQ measurement task. This approach prevents an automatic update of the data source. We did not require contact to the customer support and the install documentation and user guide was up-to-date and very intuitive to use. 

\subsubsection{Talend Open Studio for Data Quality}
% \footnote{http://www.talend.com (June 2019)}
The company Talend offers two DQ products: Talend Open Studio (TOS) for Data Quality (a free version) and Talend Data Management Platform (requires subscription). 
Gartner upgraded Talend in their Magic Quadrant of Data Quality Tools from being ``visionary'' in 2016 to ``leader'' in 2017~\cite{GartnerDQTools2016, GartnerDQTools2017}. Talend Open Studio for Data Quality is one of the most frequently cited DQ tools that we discovered in our systematic search: it was found through Springer Link and GitHub\footnote{\url{https://github.com/Talend/tdq-studio-se} (June 2019)} and was already previously investigated in~\cite{Pushkarev_DQSurvey_2010,Pulla_DQSurvey_2016,GaoEtAl_BigDataQA_2016}. 
Both products (Open Studio and Enterprise) offer good support for Big Data analysis like Spark or Hadoop and a variety of data profiling and cleansing functionalities. We evaluated version 6.5.1 of TOS for Data Quality, which can definitely keep up with several commercial DQ tools (which require a fee) in terms of data profiling capabilities, business rule management, and UI experience. However, the free version does not support DQ monitoring capabilities, which is an exclusive feature of the Enterprise edition. It was not possible to receive a free trial of the Talend Data Management Platform, because according to our customer contact, it is unlikely that someone would purchase the Enterprise edition because of this feature. 

\subsubsection{SAS Data Quality}
% \footnote{http://www.sas.com (June 2019)}
The US company SAS (Statistical Analysis System) offers three commercial DQ products: SAS Data Management, SAS Data Quality, and SAS Data Quality Desktop~\cite{GartnerDQTools2019}. Our SAS customer support offered us SAS Base 9.4 with SAS Data Quality Desktop 2.7 in a 60-days trial for the evaluation. We found SAS Data Quality through the Gartner survey~\cite{GartnerDQTools2017} and Google search. 
Since the traditional focus of SAS is on data analysis, their DQ product is based on the acquired company DataFlux. The product ``dfPower'' by DataFlux has previously been surveyed by Barateiro and Galhardas~\cite{Barateiro_DQSurvey_2005} and is mentioned in~\cite{Maletic_2009}, which was discovered through our systematic search. In our evaluation, we did not find powerful machine learning (ML) capabilities (as core strength of SAS) in DQ measurement, which was also mentioned in~\cite{GartnerDQTools2017}. According to our customer contact and also mentioned in~\cite{GartnerDQTools2019}, SAS' overall strategic focus is on migrating all product lines into the cloud-based SAS Viya platform to increase the usability and to better integrate ML and DQ. 
In the evaluated tool SAS Data Quality Desktop 2.7, we found that the overall usability was below the average when compared to other DQ tools. The customer support was friendly, but hardly any question could be answered directly. 

\subsubsection{Data Quality Solutions dedicated to SAP}
SAP (German abbreviation for ``Systeme, Anwendungen und Produkte in der Datenverarbeitung'', i.e., ``Systems, Applications and Products in Data Processing''), is a worldwide operating company for enterprise application software with headquarters in Germany. 
Since SAP is market leader in the data processing domain, there are several DQ tools that are specifically built to operate on top of an existing SAP installation. During this survey, we had no access to such an installation, and thus, were not able to include those tools in our evaluation. However, due to the practical relevance of DQ measurement in SAP, we describe the most relevant tools dedicated to SAP, which we found through our systematic search.

\paragraph{SAP Information Steward.}
SAP Information Steward was found through our systematic search and previously mentioned in~\cite{GartnerDQTools2019,Abedjan_2015,Abedjan2019}. According to the documentation, the tool offers different data profiling functionalities (like simple statistics, histograms, data types, and dependencies), allows to define and execute business rules, as well as to monitor DQ with scorecards. Its strength are the wide range of out-of-the-box functions for specific domains like customers, supply chains, and products, however, customers often state that the costs for the product are too high and the interface needs some modernization for business users~\cite{GartnerDQTools2019}. 

\paragraph{Data Quality Solution by ISO Professional Services.}
The German company ISO Professional Services offers a data governance solution, which is implemented directly in SAP and reuses user-defined business rules from the SAP environment. A few years ago, ISO acquired the company Scarus Software GmbH with the DQ tool DataGovernanceSuite, which was discovered through our search and was previously evaluated in~\cite{Fraunhofer2012}. The Scarus Data Quality (SDQ) Server constitutes the core DQ component by ISO, which has a separate memory but no DB. SDQ interoperates with SAP transparently, by offering functions like data profiling, duplicate detection, and address validation, which are directly executed within SAP. 
In contrast to its competing product SAP Information Steward, which aims at large enterprises, the tool by ISO is optimized for small to medium-sized companies. Reference customers of this size preferred the tool by ISO  Professional Services due to its adjusted functional scope and cheaper pricing. 

\paragraph{dspCompose by BackOffice Associates GmbH.}
The German company BackOffice Associates GmbH offers a DQ suite prefixed with ``dsp'' (data stewardship platform), which is dedicated to master data management. Their primary DQ products are dspMonitor (for data profiling, monitoring, and DQ checks), which is a competing product to SAP Information Steward, and dspCompose (for data cleansing and DQ workflow management), which acts as add-on for dspMonitor or SAP Information Steward. Further DQ related products are dspMigrate, an end-to-end data migration tool, dspConduct, a SAP MDE tool, and dspArchive for data achiving in SAP environments. 
Although BackOffice Associates offer their DQ products to customers without SAP, they developed a strong SAP focus in recent years. According to our customer contact, they leverage the greatest potential in offering dspCompose in combination with SAP. 

\subsection{Comparison of Data Profiling, DQ Measurement, and Monitoring Capabilities}
In this Section, we investigate the DQ tools with regard to our catalog of requirements from Table~\ref{tab:reqCatalog}. For each requirement, three ratings are possible: (\checkmark) the requirement is fulfilled, ($-$) the requirement is not fulfilled, or ($p$) the requirement is partially fulfilled. The coverage of each requirement is described in textual form with a focus on the justification of partial fulfillments. 

\begin{table}[ht]
    \caption{Data Profiling Capabilities}
    \label{tab:survey_dp}
    \footnotesize
    \begin{center}
    \begin{tabular}{|ll|c|c|c|c|c|c|c|c|c|c|c|}
    \cline{3-13}
    \multicolumn{2}{c|}{}
         &\rotatebox{90}{Aggregate Profiler}&\rotatebox{90}{Ataccama ONE}&\rotatebox{90}{DataCleaner}&\rotatebox{90}{Datamartist}&\rotatebox{90}{Experian Pandora}&\rotatebox{90}{Informatica DQ}&\rotatebox{90}{InfoZoom \& IZDQ}&\rotatebox{90}{OpenRefine \& MetricDoc~}&\rotatebox{90}{Oracle EDQ}&\rotatebox{90}{SAS Data Quality}&\rotatebox{90}{Talend Open Studio}\\\hline\hline
         %%% SC - Cardinalities
         1&Number of rows&\checkmark&\checkmark&\checkmark&\checkmark&\checkmark&\checkmark&\checkmark&\checkmark&\checkmark&\checkmark&\checkmark\\
         2&Number of nulls&\checkmark&\checkmark&\checkmark&\checkmark&\checkmark&\checkmark&\checkmark&\checkmark&\checkmark&\checkmark&\checkmark\\
         3&Percentage of nulls&$-$&\checkmark&$-$&\checkmark&\checkmark&\checkmark&\checkmark&\checkmark&\checkmark&\checkmark&\checkmark\\
         4&Number of distinct values&\checkmark&\checkmark&\checkmark&\checkmark&\checkmark&\checkmark&\checkmark&\checkmark&\checkmark&\checkmark&\checkmark\\
         5&Percentage of distinct values&$-$&\checkmark&$-$&\checkmark&\checkmark&\checkmark&\checkmark&$-$&$-$&\checkmark&\checkmark\\\hline
         
         %%% SC - Value distributions
         6&Frequency histograms&$-$&$-$&$p$&$p$&$p$&$p$&$p$&$p$&$p$&$-$&$p$\\
         7&Minimum and maximum values&\checkmark&\checkmark&\checkmark&\checkmark&\checkmark&\checkmark&\checkmark&$-$&\checkmark&\checkmark&\checkmark\\
         8&Constancy&$-$&\checkmark&$-$&\checkmark&\checkmark&\checkmark&\checkmark&$-$&\checkmark&\checkmark&\checkmark\\
         9&Quartiles&\checkmark&$p$&\checkmark&$-$&$-$&$-$&$p$&$-$&$-$&$p$&\checkmark\\
         10&Distribution of first digit&$-$&$-$&$-$&$-$&$-$&$-$&$-$&$-$&$-$&$-$&\checkmark\\\hline
         
         %%% SC - Data types, patterns, and domains
         11&Basic types&$-$&\checkmark&$p$&\checkmark&\checkmark&\checkmark&\checkmark&\checkmark&\checkmark&\checkmark&$-$\\
         12&DBMS-specific data type&\checkmark&$-$&\checkmark&$-$&$-$&\checkmark&$-$&$-$&$-$&\checkmark&\checkmark\\
         13&Value length&\checkmark&\checkmark&\checkmark&$-$&\checkmark&$p$&$p$&$p$&$p$&$p$&$p$\\
         14&Number of digits&\checkmark&$-$&\checkmark&$-$&\checkmark&\checkmark&\checkmark&$-$&\checkmark&$-$&$-$\\
         15&Number of decimals&\checkmark&$-$&$-$&$-$&\checkmark&\checkmark&\checkmark&$-$&$-$&$-$&$-$\\
         16&Histogram of value patterns&$-$&$-$&\checkmark&\checkmark&\checkmark&\checkmark&\checkmark&$-$&\checkmark&$p$&\checkmark\\
         17&Generic semantic data type&$-$&$-$&\checkmark&$-$&\checkmark&\checkmark&$-$&$-$&\checkmark&\checkmark&\checkmark\\
         18&Semantic domain&$-$&$-$&\checkmark&$-$&\checkmark&\checkmark&$-$&$-$&\checkmark&\checkmark&\checkmark\\\hline
         
         %%% Dependencies
         19&UCCs (key discovery)&$-$&$-$&$-$&$-$&$p$&\checkmark&$-$&$-$&$p$&$p$&$-$\\
         20&Relaxed UCCs&$-$&$-$&$-$&$-$&$p$&\checkmark&$-$&$-$&$-$&$-$&$-$\\
         21&INDs (foreign key discovery)&$-$&$-$&$-$&$-$&\checkmark&$p$&$-$&$-$&$-$&$p$&$-$\\
         22&Relaxed INDs&$-$&$-$&$-$&$-$&\checkmark&$p$&$-$&$-$&$-$&$p$&$-$\\
         23&FDs&$-$&$-$&$-$&$-$&\checkmark&\checkmark&$-$&$-$&$-$&$-$&$p$\\
         24&Relaxed FDs&$-$&$-$&$-$&$-$&\checkmark&\checkmark&$-$&$-$&$-$&$-$&$p$\\\hline
         
         %%% Advanced data profiling
         25&Correlation analysis&\checkmark&$-$&$-$&$-$&$-$&$-$&$-$&$-$&$-$&$-$&$p$\\
         26&Association rule mining&$-$&$-$&$-$&$-$&$-$&$-$&$-$&$-$&$-$&$-$&$-$\\
         27&Cluster analysis&$p$&$-$&$-$&$p$&$-$&$p$&$-$&$p$&$p$&$p$&$-$\\
         28&Outlier detection&$p$&$p$&$-$&$p$&\checkmark&$p$&$p$&$-$&$-$&$p$&$-$\\
         29&Exact duplicate detection&$p$&$-$&\checkmark&$p$&\checkmark&\checkmark&\checkmark&$p$&\checkmark&\checkmark&\checkmark\\
         30&Relaxed duplicate detection&$p$&$-$&\checkmark&$-$&\checkmark&\checkmark&\checkmark&$p$&\checkmark&\checkmark&\checkmark\\\hline
\end{tabular}
\end{center}
\end{table}

\subsubsection{Data Profiling Capabilities}
Table~\ref{tab:survey_dp} shows the fulfillment of data profiling capabilities for each tool. We excluded Apache Griffin and MobyDQ from this table, because both tools do not offer any data profiling functionality. It can be summarized that basic single-column data profiling like cardinalities (DP 1--5) are covered by most tools, but more sophisticated functionalities, like dependency discovery and multi-column profiling, are offered only in single cases. 

\paragraph{Single Column - Cardinalities.} While simple counts of values (i.e., cardinalities), like the number of rows, \texttt{null} values, or distinct values are covered by all DQ tools that support data profiling in general, the major distinction is an out-of-the-box availability of percentage values. The percentage of \texttt{null} values (DP-3) or distinct values (DP-5) is not supported by all investigated tools. In addition, the test case results reveal different precision for the calculation of the percentages. For example, the percentage of \texttt{null} values in column \texttt{Supplier}.\texttt{Fax} was 55~\% with Datamartist, 55.2~\% with Oracle EDQ and SAS DataFlux, and 55.17~\% in all other tools. 
The test case for DP-5 yielded 23~\% with Datamartist, 23.07~\% with Informatica and 23.08~\% with the other tools.

\paragraph{Single Column - Value Distributions.} Value distributions can be described as cardinalities of value groups~\cite{Abedjan_2015}. 
While histograms to visualize value distributions are available in most tools in the form of equi-width histograms (which ``span value ranges of same length''~\cite{Abedjan_2015}), we did not find any tool that supports equi-depth or equi-height histograms (where each bucket represents ``the same number of value occurrences''~\cite{Abedjan_2015}). Thus, we rated all tools that support histograms with ``partially'' for DP-6. 
Ataccama allows frequency analysis but no visualization with histograms, and Aggregate Profiler visualizes the distributions only in form of a pie chart. 
The majority of tools also support minimum and maximum values (DP-7), as well as constancy (DP-8), which is defined as ``the ratio of the frequency of the most frequent value (possibly a pre-defined default value) and the overall number of values''~\cite{Abedjan_2015}. 
% most frequent value and the total number of values~\cite{Abedjan_2015}. 
``Benford's law'' (DP-10), which is particularly interesting in the area of fraud detection, was only available in Talend OS. 

Quantiles are a statistical measure to divide a value distribution into equidistant percentage points~\cite{Sheskin_2003}. The most common type of quantiles, which we observed in our study, are ``quartiles'' (DP-9), where the value distribution is divided by three points into four blocks. The division points are a multiple of 25~\%, denoted as lower quartile or Q1 (25~\%), median or Q2 (50~\%), and upper quartile or Q3 (75~\%), respectively.
Other examples for quantiles are ``percentiles'', which divide the distribution into 100 blocks (i.e., each block comprises a proportion of 1~\%), or ``deciles'', which divide the distribution into 10 blocks of each 10~\% value distribution~\cite{Sheskin_2003}. While only three tools explicitly support quartiles, we discovered the availability of other types of quantiles too (in our survey rated as $p$). Table~\ref{tab:dp_test_cases2} shows the results for the DP-9 test case where quartiles or other types of quantiles are calculated for the column \texttt{OrderItem.UnitPrice}. 

\begin{table}[ht]
    \caption{Data Profiling - Test Case Quartiles}
    \label{tab:dp_test_cases2}
    \footnotesize
    \begin{center}
    \begin{tabular}{lllll}\hline
    &Q1 (25~\%)&Q2 (50~\%) &Q3 (75~\%) &Type of Quantile\\\hline
    Aggregate Profiler&12&18.4&32&Quartiles (4 blocks)\\
    DataCleaner~&12&18.4&32&Quartiles (4 blocks)\\
    Talend Open Studio&12&18.4&32&Quartiles (4 blocks)\\
    SAS Data Quality&12.9375&19.475&33.4375&Demi-deciles (20 blocks)\\
    InfoZoom \& IZDQ&12 (25.48 \%)&18.4 (50.63~\%)&32 (75.36~\%)&Inverse function (2.155 blocks)\\\hline
    Ataccama ONE&\multicolumn{3}{p{6.5cm}}{0~\%: 2, 10~\%: 7.45, 20~\%: 10, 30~\%: 13.25, 40~\%: 16, 50~\%: 18.4, 60~\%: 21.5, 70~\%: 30, 80~\%: 35.1, 90~\%: 46, 100~\%: 263.5}&Deciles (10 blocks)\\
         \hline
\end{tabular}
\end{center}
\end{table}

Ataccama ONE supports deciles, which are displayed extra in the last row since they cannot be directly mapped to quartiles. Although SAS Data Quality provides 20 blocks, that is, demi-deciles, the functionality is described in the SAS UI as ``percentiles'', which would refer to the 100-partitions quantiles. In Table~\ref{tab:dp_test_cases2}, we picked only the values for the quartile blocks out of the 20 blocks in total. 
In InfoZoom, the inverse function to quantiles is chosen: instead of merging values into blocks, the percentage value of the distribution is display for each value (denoted as ``Cumulative Distribution Function''~\cite{DasuJohnson_2003}), leading to a total of 2.155 blocks, where each block contains exactly one value.  Table~\ref{tab:dp_test_cases2} displays the percentage value of the distribution that refers to Q1, Q2, and Q3, respectively. 
It can be summarized that the determination of quantiles is interpreted differently in the single DQ tools with respect to the notation (``Q1'' vs. ``lower quartile'' vs. 25~\%) as well as the type of quantile. 

\paragraph{Single Column - Patterns, Data Types, and Domains.} In this category, the support of the different requirements varies widely and there is definitive potential for improvement with respect to out-of-the-box pattern and domain discovery. 
Even the discovery of basic types (DP-11) is not always supported. For example, DataCleaner recognizes the difference between string, boolean, and number and uses this information for further internal processing, but does not explicitly display it per attribute. 
While the test cases for the DBMS-specific data types (DP-12) yielded uniform results (``varchar'' for \texttt{ProductName}, ``decimal'' for \texttt{UnitPrice}, and ``bit'' for \texttt{isDiscontinued}), the variety in terminology and classification for the basic types is outlined in Table~\ref{tab:dp_test_cases_11}. 
In SAS, we had problems to access a table containing the ``decimal'' data type and thus, converted \texttt{Product.UnitPrice} to ``long''. ``Alphanumeric'' in Experian Pandora is abbreviated with ``Alphanum.''. 

\begin{table}[ht]
    \caption{Data Profiling - Test Cases Basic Types}
    \label{tab:dp_test_cases_11}
    \footnotesize
    \begin{center}
    \begin{tabular}{|l|c|c|c|c|c|c|c|c|}
    \cline{2-9}
    \multicolumn{1}{c|}{}
         &A-ONE&DM&EP&IDQ&IZDQ&OR&O-EDQ&SAS\\\hline
         \texttt{ProductName} &String&Text&Alphanum.&String(32)&String&String&Text&String\\
         \texttt{UnitPrice} &String&Number&Decimal&Decimal(5,2)&\#\#\#\#.\#\#&String/numeric&Numeric&Long\\
         \texttt{isDiscontinued} &Integer&Number&Integer&Integer(1)&\#\#\#\#&Numeric&Numeric&Bit\\
         \hline
\end{tabular}
\end{center}
\end{table}

For the measurement of the value length (DP-13), the minimum (min.) and maximum (max.) values are usually provided, but not always an average (avg.) value length. The median (med.) value length is only provided by Ataccama ONE. We rated this requirement as fulfilled if at least the minimum, maximum, and average value length were provided, considering the median as optional. Table~\ref{tab:dp_test_cases} shows the exact results delivered by the single tools, which justifies the fulfillment ratings and indicates differences in the accuracy of the average values. InfoZoom provides only the maximum value length, while SAS and Talend OS restrict this feature to string values. 

\begin{table}[ht]
    \caption{Data Profiling - Test Case Value Length}
    \label{tab:dp_test_cases}
    \footnotesize
    \begin{center}
    \begin{tabular}{|l|c|c|c|c|c|c|c|c|c|c|}
    \cline{2-11}
    \multicolumn{1}{c|}{}
         &AP&A-ONE&DC&EP&IDQ&IZDQ&OR&O-EDQ&SAS&TOS\\\hline
         \texttt{ProductName} (min.) &4&4&4&4&4&$-$&4&4&4&4\\
         \texttt{ProductName} (max.) &32&32&32&32&32&32&32&32&32&32\\
         \texttt{ProductName} (avg.) &16.269&16.27&16.269&16.32&$-$&$-$&16.269&$-$&$-$&16.32\\
         \texttt{ProductName} (med.) &$-$&15&$-$&$-$&$-$&$-$&$-$&$-$&$-$&$-$\\
         \hline
\end{tabular}
\end{center}
\end{table}

For the number of digits and decimals, the DQ tools usually use the values documented by the DBMS, e.g., 12 digits and 2 decimals for attribute \texttt{UnitPrice} in table \texttt{Product}, compared to maximum 5 digits and 2 decimals in the real data. Value patterns and their visualization as a histogram (DP-16) is supported by most DQ tools. SAS supports pie charts only. 

Generic semantic data types (DP-17), such as code, indicators, date/time, quantity, or identifier are also denoted as ``data class''~\cite{Abedjan_2015} and are defined by generic patterns. A semantic domain (DP-18), ``such as a credit card, first name, city, [or] phenotype''~\cite{Abedjan_2015}, is more concrete than a generic semantic data type and usually associated with a specific application context. The DQ tools that fulfill these requirements offer a number of patterns, which are associated with the respective generic data type or semantic domain. By applying these patterns to the data values, it could be verified to which extent an attribute contains values that are of a specific type. Thus, the two requirements DP-17 and DP-18 are usually not distinguished within the DQ tools we evaluated. 
The number of available patterns varies between approximately 10-50 patterns (Pandora, DataCleaner, SAS), 50-100 patterns (Talend), and 100-300 (Informatica, Oracle). 
While most tools display the matching patterns per attribute (e.g., \texttt{Product.UnitPrice} conforms to 98.72~\% to the domain ``Geocode\_Longitude'' using Informatica DQ), SAS displays the matching attribute per pattern (e.g., ``Country'' matches to 100~\% the attribute \texttt{Customer.Country}). 
Talend OS is the only tool that displays the matching rows instead of the percentage of matching rows per attribute. 
For Ataccama ONE, we rated DP-17 and DP-18 as partially fulfilled, since specific attributes are classified (e.g., \texttt{Customer.FirstName} as ``first name''), but those terms are part of the Ataccama business glossary, which we were unable to access during our evaluation and therefore, had no further information about its origin.

\paragraph{Dependencies.}
The dependency section has the lowest coverage of the data profiling category and is best supported by Experian Pandora and Informatica DQ (in Developer edition only). 
Although we introduce each concept briefly, we refer to~\cite{Abedjan2019} for details about dependency discovery and their implementation. In the following, $\mathcal{R}$ denotes a relational schema (defining a set of attributes) with $r$ being an instance of $\mathcal{R}$ (defining a set of records). Sets of attributes are denoted by $\alpha$ and $\beta$.

A unique column combination (UCC) is an attribute set $\alpha \subseteq \mathcal{R}$ whose projection contains no duplicate entries in $r$~\cite{Abedjan2019}. In other words, a UCC is a (possibly composite) candidate key that functionally determines $\mathcal{R}$. 
While Experian Pandora allows the detection of single column keys (thus $p$), Informatica DQ offers full UCC detection. Both tools allow the user to set a threshold for relaxed UCC detection (DP-20) and to identify violating records via drill-down. With Informatica DQ, we discovered five UCCs in table \texttt{Order} of our test DB, using a threshold of 98~\%: \texttt{Id} (100~\%), \texttt{OrderNumber} (100~\%), \texttt{OrderDate} + \texttt{TotalAmount} (100~\%), \texttt{CustomerId} + \texttt{TotalAmount} (99.88~\%), and \texttt{CustomerId} + \texttt{OrderDate} (99.16~\%). With Experian Pandora, only the two single column keys \texttt{Id} (100~\%) and \texttt{OrderNumber} (100~\%) were detected. SAS Data Quality indicates 100~\% unique attributes as primary key candidates. 

An inclusion dependency (IND) over the relational schemata $R_{i}$ and $R_{j}$ states that all values in attribute set $\alpha$ also occur in $\beta$, that is $R_{i}[\alpha ] \subseteq R_{i}[\beta ]$~\cite{Abedjan2019}. 
The detection of INDs (DP-21 and DP-22), also referred to as foreign key discovery, is not widely supported. The best automation for this requirement delivers Experian Pandora, where initially the primary keys (UCCs) and foreign key relations are inferred, and based on this information, INDs are displayed graphically as a Venn diagram. In addition, it is possible to drill down to records that violate those INDs in a spread sheet. Informatica DQ and SAS Data Quality support IND discovery only partially, since it is required that the user selects the respective primary key (UCCs) and assigns it to possible foreign key candidates that are then tested for compliance. 
DataCleaner can only be used to check if two tables can possibly be joined, without information on the respective columns or the join quality (i.e., violating rows).

A functional dependency (FD) $\alpha\rightarrow\beta$ asserts that all pairs of records with the same attribute values in $\alpha$ must also have the same attribute values in $\beta$. Thus, the $\alpha$-values functionally determine the $\beta$-values~\cite{Codd_1970}. Again, we used table \texttt{Order} to verify exact (DP-23) and relaxed (DP-24) FD detection. With Experian Pandora and Informatica DQ we found in total eight exact FDs: \{\texttt{Id}\}~$\rightarrow$~\{\texttt{OrderDate}, \texttt{OrderNumber}, \texttt{CustomerId}, \texttt{TotalAmount}\} and \{\texttt{OrderNumber}\}~$\rightarrow$~\{\texttt{Id}, \texttt{OrderDate}, \texttt{CustomerId}, \texttt{TotalAmount}\} and two more FDs when relaxing the threshold to 93~\%: \{\texttt{TotalAmount}\}~$\rightarrow$~\{\texttt{CustomerId}, \texttt{OrderDate}\}. 
Talend OS fulfills FD discovery only partially, because it requires user interaction to specify the attribute sets $\alpha$ and $\beta$, given that the number and type of columns are equal. Although specific FDs can be tested with this functionality (e.g., to which extent \texttt{TotalAmount} determines \texttt{CustomerId}), we do not perceive it as true automated FD discovery. E.g., when performing the test case and specifying all attributes of \texttt{Order} as $\alpha$ and $\beta$ respectively, the result are five FDs, where each attribute is discovered to functionally determine itself. This case should be ideally excluded during the detection. 
All three tools printed the identified FDs in table-format, one row for each attribute pair along with the match percentage, but with  slightly differing terminology. Thus $\alpha$ (the left side) is denoted as ``A column set'', ``identity column'' or ``determinant column'', and $\beta$ (the right side) is denoted as ``B column set'', ``identified columns'' or ``dependent column''. For relaxed FDs, Experian displayed the violating rows with a count (50 in this case), Informatica listed the respective rows, and Talend did not provide violating rows at all. 

\paragraph{Advanced Multi-Column Profiling.} Apart from duplicate detection, which is a widely supported feature, advanced multi-column features are rarely supported satisfactorily. No single tool offers association rule mining (DP-26) as mentioned in~\cite{Abedjan_2015}. Note that we specifically tested the DQ tools described in Section~\ref{sec:tool_descr}, and did not consider related tools that are often installed together. For example, SAS Enterprise Guide, which was shipped with our DQ installation, is dedicated to data analysis and therefore provides a rich function palette that overlaps with the multi-column profiling section, e.g., a selection of correlation coefficients, hierarchical and k-means clustering. Since the aim of this survey is to investigate DQ tools, we did not consider such related tools. 

Correlations (DP-25) are a statistical measure between 1.0 and -1.0 to indicate the relationship between two numerical attributes~\cite{Abedjan_2015,Sheskin_2003}. The most commonly used coefficients are Pearson correlation coefficient, or the rank-based Spearman's or Kendall's tau correlation coefficients~\cite{Sheskin_2003}. In our survey, only Aggregate Profiler is able to compute Pearson correlations.
However, our test case for DP-25 (Pearson correlation between \texttt{OrderItem.UnitPrice} and \texttt{OrderItem.Quantity}) yielded -0.045350608 with Aggregate Profiler, which did not conform to our cross-check using SAS Enterprise Guide (0.00737) and the Python package numpy (0.00736647). 
Talend distinguishes between ``numerical'', ``time'', and ``nominal'' correlation analysis and displays the respective correlations in bubble charts. We rated this as partial fulfillment $p$, since no correlation coefficient is calculated and the calculation is restricted to single columns with specific data types, thus, it is not possible to calculate the correlation between two interval data types. 

During our investigation, we found that the concepts of clustering (DP-27), outlier detection (DP-28), and duplicate detection (DP-29 and DP-30) are not always clearly distinguishable in practice. Also Abedjan et al.~\cite{Abedjan_2015} state that clustering can be either used to detect outliers in a single column, or to detect similar or duplicate records within a table. Thus, we describe the three concepts briefly along with the condition we applied to verify (partial) fulfillment of the respective requirement.

%% Clustering
Clustering (DP-27) is a type of unsupervised machine learning, where data items (e.g., records or attribute values) are classified into groups (clusters)~\cite{Jain_2000}. A comprehensive overview on existing clustering algorithms is provided by Jain et al.~\cite{Jain_2000}. In some DQ tools, clustering is only available in the frame of duplicate detection. 
For example, in OpenRefine clustering is used to detect duplicate string values (cf.~\cite{OpenRefinceClustering}), in Informatica DQ the grouped duplicates are referred to as ``clusters'', SAS Data Quality requires a ``clustering'' component to group records based on their match codes~\cite{SAS_doku}; and Oracle EDQ uses clustering as preprocessing step of the matching component to increase runtime efficiency by preventing unnecessary comparisons between records~\cite{Oracle_doku}.
To completely fulfill requirement DP-27, we presumed the availability of one of the common clustering algorithms (like k-means or hierarchical clustering) as an independent function. Datamartist supports k-means clustering and allows to select the number of clusters k from 5 pre-defined values (5, 10, 25, 50, 100) and to restrict the observed value range. 
Aggregate Profiler supports k-means clustering without any modification possibility (e.g., choose k) as well as a second type of clustering for numeric values, where the number of clusters can be defined. No further information about this clustering algorithm is provided. No tool offers hierarchical clustering or other partitional clustering algorithms except k-means, for example, graph theoretic approaches or expectation maximization~\cite{Jain_2000}. 

% Outlier detection
Outlier detection deals with data points that are considered abnormalities, deviants, or discordant when compared to the remaining data~\cite{Aggarwal_2017}. A comprehensive overview on different algorithms to detect outliers is provided in~\cite{Aggarwal_2017}. Our investigation showed that outlier detection is implemented in the tools very differently, and compared to the current state of research, only simple methods are used. We have not found a tool that supports multivariate outlier detection or one of the more sophisticated approaches like z-score, linear regression models, or probabilistic models as mentioned in~\cite{Aggarwal_2017}. 
In the following, we describe the implementation of outlier detection and the result that our test case yielded to ``find very high values'' in \texttt{Order.TotalAmount}:

\begin{itemize}
    \item Aggregate Profiler, Ataccama ONE, Datamartist, and InfoZoom provide outlier detection for numerical values only visually, either in a quantile plot (Ataccama), in a bar chart (Datamartist) or in form of a box plot. Aggregate Profiler and Ataccama ONE do not allow drill-down to the actual outlying values and in InfoZoom the visualization of the single values in the plot are not readable. In all three tools, it is not possible to modify the plot settings or to get details about the used settings. The bar chart in Datamartist is based on k-means clustering with the same modification options as described in the previous paragraph. When using the standard settings (100 bars), one outlier is detected for our test case of finding ``very high values'' in column \texttt{Product.UnitPrices}: 17250.0. This extreme value is detected by all tools correctly, although other methods yield more outlying values.
    \item Experian Pandora offers a number of different types of outlier checks, where some require one of the two parameters that can be specified by a user: ``Rarity Threshold'' (default: 1000) and ``Standard Deviation Tolerance'' (default: 3.3)~\cite{Experian_doku}. The rarity threshold is used to detect rare values, which occur less frequently than one time in <threshold> and is used for the checks ``rare values'', ``is a key'', and ``unusually missing values''~\cite{Experian_doku}. The standard deviation tolerance specifies the number of standard deviations that is tolerated for a value to be apart from the norm. It is used for low/high amounts, short/long values, rare/frequent values, and rare formats~\cite{Experian_doku}. 
    By using the standard settings, we found 18 outlying values for our test case (17250.0, 16321.9, 15810.0, 12281.2, 11493.2, 11490.7, 11380.0, 11283.2, 10835.24, 10741.6, 10588.5, 10495.6, 10191.7, 10164.8, 8902.5, 8891.0, 8623.45, 8267.4).
    \item Informatica DQ distinguishes between ``pattern outliers'', which refer to unusual patterns in the data, and ``value frequency outliers''~\cite{Informatica_doku_profiler}, where values with unusual occurring frequency are displayed. With this functionality, it was not possible to perform our test case, because the characteristic of being an outlier depends on the  frequency instead of the actual value.
    \item SAS provides an ``outliers'' tab for columns of different data types, where a fixed number of five minimum and maximum values are outlined without any modification possibility. For our test case, the following maximum values have been detected: 17250.0, 16321.0, 15810.0, 12281.2, and 11493.2, which correspond to the five highest results detected with Pandora.
\end{itemize}

%% Duplicate detection
Duplicate detection ``aims to identify records [...] that refer to the same real-world entity"~\cite{DuplicateDetection_2009}.
It is a widely researched field, which is also referred to as record matching, record linkage, data merging, or redundancy detection~\cite{DuplicateDetection_2009}. In contrast to clustering and outlier detection, the understanding and implementation of duplicate detection is very similar across all tools we investigated. 
In principle, the user (1) selects the columns that should be considered for comparison, (2) optionally applies a transformation to those columns (e.g., pruning a string to the first three characters), and finally (3) selects an appropriate distance function and algorithm. 
The major difference in the implementations is the selection of distance functions for the attribute values. The following distances are supported:

\begin{itemize}
    \item Aggregate Profiler: exact match, similar-any word (if any word is similar for this column), similar-all words (if all words are similar for this column), begin char match, and end char match~\cite{AP_doku}. No information about the used similarity function was provided.
    \item DataCleaner: n-grams, first 5, last 5, sorted acronym, Metaphone, common integer, Fingerprint, near integer (for pre-selection phase); exact, is empty, normalized affine gap, and cosine similarity (for scoring phase). The two phases are explained in the following paragraph.
    \item Experian Pandora: Edit distance, exact, exact (ignore cases), Jaro distance, Jaro-Winkler distance, regular expression, and Soundex.
    \item Informatica DQ: Bigram, Edit, Hamming, reverse Hamming, and Jaro distance. 
    \item InfoZoom: Soundex and Cologne phonetics.
    \item OpenRefine: Fingerprint, n-gram Fingerprint, Metaphone3, or Cologne phonetics (with key collision method); Levenshtein or PPM (with nearest neighbor method)~cf.~\cite{OpenRefinceClustering} for details.
    \item Oracle EDQ: (transformations) absolute value, first/last n characters/words, lower case, Metaphone, normalize whitespace, round, Soundex. 
    \item Talend OS: exact, exact (ignore case), Soundex, Soundex FR, Levenshtein, Metaphone, Double Metaphone, Jaro, Fingerprint key, Jaro-Winkler, q-grams, Hamming, and custom. 
\end{itemize}

SAS Data Quality does not offer string distances, but matches based on match codes~\cite{SAS_doku}, which are generated based on an input variable, a ``definition'' (type of transformation for the input variable) and a ``sensitivy'' (threshold), where records with the same match codes are then grouped together into the same cluster. The list of match definitions depends on the used Quality Knowledge Base (QBK).
Talend OS offers two different algorithms to define the record merge strategy: simple VSR Matcher (default) or T-Swoosh. We refer to the documentation~\cite{Talend_doku} for details. 
%OpenRefine offers duplicate detection on attribute-level only, which is why we rated it as partial fulfillment. 

DataCleaner implements a ML-based approach that distinguishes between two modes: untrained detection (considered experimental) and a training mode plus duplicate detection using the trained ML model~\cite{Quadient_doku}. The training mode is divided into three phases: (1) pre-selection, (2) scoring using a random forest classfier and the distance functions mentined above, and (3) the outcome, which highlights duplicate pairs with a probability between 0 and 1.

Despite the fact that duplicate detection is typically attributed towards data cleansing (data integration or data matching) and not considered to be part of data profiling in the implementations, most DQ tools allow this functionality to be used also for detection purposes. 
We rated Datamartist and Aggreate Profiler as supporting this requirement partially since the function is dedicated to direct cleansing (deletion or replacement of records) and because of the very limited configuration options compared to all other tools. Datamartist does not support DP-30 at all.

\subsubsection{Data Quality Measurement Capabilities}
\label{sec:eval_DQM}
Table~\ref{tab:survey_dqm} summarizes the fulfillment of the DQM category, where the first part is dedicated to DQ dimensions, and the second one to business rules. 
%% Accuracy
A metric to measure the \textit{accuracy} on table-level is only provided by Apache Griffin, where the user needs to select a source and a target table and accuracy is calculated according to $A_{tab}=\frac{|r_a|}{|r|}*100\%$, where $|r|$ is the total number of records in the source table, and $|r_a|$ the number of (accurate) records in the target table that can be directly matched to a record in the source table~\cite{Griffin_doku}. This metric is based on the definition in~\cite{DAMA_dimensions} and corresponds to the accuracy metric proposed by Redman~\cite{Redman_2005}.

\begin{table}[ht]
    \centering
    \caption{Data Quality Measurement Capabilities}
    \label{tab:survey_dqm}~
    \footnotesize
    \begin{tabular}{|ll|c|c|c|c|c|c|c|c|c|c|c|c|c|}
    \cline{3-15}
    \multicolumn{2}{c|}{}
         &\rotatebox{90}{Aggregate Profiler}&\rotatebox{90}{Apache Griffin}&\rotatebox{90}{Ataccama ONE}&\rotatebox{90}{DataCleaner}&\rotatebox{90}{Datamartist}&\rotatebox{90}{Experian Pandora}&\rotatebox{90}{Informatica DQ}&\rotatebox{90}{InfoZoom \& IZDQ}&\rotatebox{90}{MobyDQ}&\rotatebox{90}{OpenRefine \& MetricDoc~}&\rotatebox{90}{Oracle EDQ}&\rotatebox{90}{SAS Data Quality}&\rotatebox{90}{Talend Open Studio}\\\hline\hline
         31&Accuracy metric&$-$&\checkmark&$-$&$-$&$-$&$-$&$-$&$-$&$-$&$-$&$-$&$-$&$-$\\
         32&Completeness metric&$-$&$-$&$p$&$p$&$p$&$p$&$p$&$p$&\checkmark&$-$&$-$&$-$&$-$\\
         33&Consistency metric&$-$&$-$&$-$&$-$&$-$&$-$&$-$&$-$&$-$&$-$&$-$&$-$&$-$\\
         34&Timeliness metric&$-$&$-$&$-$&$-$&$-$&$-$&$-$&$-$&$-$&$-$&$-$&$-$&$-$\\
         35&Other DQ metrics&$-$&$-$&$-$&$p$&$p$&$p$&$p$&$-$&\checkmark&\checkmark&$-$&$p$&$p$\\\hline
         36&Creation of business rules&\checkmark&$-$&$-$&$-$&\checkmark&\checkmark&\checkmark&\checkmark&\checkmark&\checkmark&\checkmark&\checkmark&\checkmark\\
         37&General-applicable rules&$-$&$-$&$-$&$-$&$-$&\checkmark&\checkmark&$p$&$-$&$-$&$-$&\checkmark&\checkmark\\
         38&Application of business rules&\checkmark&$-$&$-$&$-$&\checkmark&\checkmark&\checkmark&\checkmark&\checkmark&\checkmark&\checkmark&\checkmark&\checkmark\\
         \hline
    \end{tabular}
\end{table}

%% Completeness
A simple metric for the \textit{completeness} on attribute-level can be calculated with $C_{att}=\frac{|v_c|}{|v|}$, where $|v|$ is the total number of values within a column and $|v_c|$ is the number of values that are not \texttt{null}. This metric is closely related to DP-3, the percentage of \texttt{null} values, which yields the missingness $M_{att}=\frac{|v_n|}{|v|}=1-C_{att}$, where $|v_n|$ is the number of \texttt{null} values. 
Ataccama, DataCleaner, Datamartist, Experian, and InfoZoom provide the completeness calculation on attribute-level, without the possibility for aggregation on higher levels. Informatica DQ allows an aggregation on table-level (as the average of all attribute-level completeness values), but not higher. 
Despite the fact that MetricDoc offers a metric that is denoted ``completeness'' in the GUI and is also described in their scientific documentation~\cite{Bors_2018}, they calculate the missingness $M_{att}$ on attribute-level and thus did not fulfill requirement DQM-32.
MobyDQ computes the completeness between two data sources (source and target) according to $C_{tab}=\frac{C_{t}-C_{s}}{C_{s}}$, where $C_{t}$ is the completeness measure of the target source and $C_{s}$ the measure from the source. Here, $C_{s}$ can be considered as the reference data set. 

%% Consistency
The \textit{consistency} dimension is mentioned in the Informatica DQ methodology~\cite{Informatica_DQ_methodology}, and SAS implements no single metric, but a set of rules that are grouped to this dimension, e.g., checks if an attribute contains numbers, non-numbers, is alphabetic, or is all lower case. We did not rate this as fulfilled since those rules were supplied by most DQ tools with generally applicable business rules (DQM-37), but usually not specifically attributed towards the consistency dimension. 

%% Timeliness
We did not find any implementation for the \textit{timeliness} dimension. However, with respect to other time-related dimensions, MetricDoc offers two different \textit{time interval metrics}, where one checks if the interval between two timestamps ``is smaller than, larger than, or equal to a given duration value''~\cite{Bors_2018}, and the second one performs outlier detection on interval length~\cite{Bors_2018}.
MobyDQ offers metrics for \textit{freshness} and \textit{latency}, but refers to those dimensions as DQ ``indicators''~\cite{MobyDQ_doku}. 
Freshness is implemented as $ts_{cur}-ts_t$, where $ts_{cur}$ is the current timestamp and $ts_t$ the last updated timestamp from the target request, and latency as $ts_s-ts_t$, where $ts_s$ is the last updated timestamp from the source request~\cite{MobyDQ_doku}. These indicators are not specifically dedicated to DQ dimensions and do not fulfill the requirement for DQ metrics to be normalized between [0,1] by Heinrich et al.~\cite{Heinrich_2018}.

%% Other metrics
With respect to other non-time-related DQ metrics (DQM-35), the \textit{uniqueness} dimension is most often implemented according to  $U_{att}=\frac{|v_u|}{|v|}$, where $|v_u|$ refers to the number of unique values within a column. Datamartist, Experian, DataCleaner, SAS Data Quality, and Talend OS implement uniqueness on attribute-level only, which corresponds to the requirement DP-5. Informatica DQ allows an aggregation on table-level but not higher. MetricDoc implements a dimension referred to as ``uniqueness'', but actually calculate the \textit{redundancy} $R_{tab}=\frac{|r_{red}|}{|r|}$ on table-level, where $|r_{red}|$ is the number of records with at least one duplicate entry in the table. The user needs to select more than one attribute within a table in order to calculate the metric and it cannot be aggregated on higher levels. In addition, MetricDoc offers metrics for the DQ dimensions \textit{validity} and \textit{plausibility}. Validity is calculated on attribute-level as $V_{att}=\frac{|v_i|}{|v|}$, where $|v_i|$ is the number of attribute values that do not comply to the column data type. Plausibility is also calculated on attribute-level as $P_{att}=\frac{|v_j|}{|v|}$, where $|v_j|$ is the number of attribute values that are outliers according to a nonrubost or robust statistical measure (mean with standard deviation, or median with interquartile range estimator, respectively)~\cite{Bors_2018}.
MobyDQ also provides a \textit{validity} indicator, which connects to one single target data source and compares the values with defined thresholds~\cite{MobyDQ_doku}. 

A number of other DQ dimensions are mentioned in documentations without having an appropriate metrics implemented: integrity by Talend~\cite{Talend_doku}, integrity, conformity, duplicates, currency, and referential integrity by Informatica~\cite{Informatica_DQ_methodology}, and validity, integrity, and structure by SAS~\cite{SAS_doku}. In SAS, DQ dimensions are used as abstraction layer to group business rules. 

%% Business rules
While the creation and application of business rules (DQM-36 and DQM-38) is supported by most DQ tools, few tools also offer predefined generally applicable business rules. 
A widely supported example are rules for address validation  (e.g., zip codes, cities, states) that tackle the prevalent problem of failed mail deliveries due to incorrect addresses, also described in~\cite{ApelEtAl_2015}.
Despite the good performance of DataCleaner in terms of  data profiling and CDQM, it does not support business rules at all.
We rated DP-37 (the availability of generally applicable rules) as only partly fulfilled for InfoZoom, because the provided rules have been created for a given demo DB schema and would need to be modified to apply to  other schemas (e.g., with other column names). 

\subsubsection{Data Quality Monitoring Capabilities} 
The results of the CDQM evaluation are shown in Table~\ref{tab:survey_cdqm}. We want to point out that for two DQ tools (Ataccama ONE and Talend OS), more advanced versions are available that support CDQM according to their vendors, but we did not investigate it. 

\begin{table}[ht]
    \caption{Data Quality Monitoring Capabilities}
    \label{tab:survey_cdqm}
    \begin{center}
    \footnotesize
    \begin{tabular}{|ll|c|c|c|c|c|c|c|c|c|c|c|c|c|}
    \cline{3-15}
    \multicolumn{2}{c|}{}
         &\rotatebox{90}{Aggregate Profiler}&\rotatebox{90}{Apache Griffin}&\rotatebox{90}{Ataccama ONE}&\rotatebox{90}{DataCleaner~}&\rotatebox{90}{Datamartist}&\rotatebox{90}{Experian Pandora}&\rotatebox{90}{Informatica DQ}&\rotatebox{90}{InfoZoom \& IZDQ}&\rotatebox{90}{MobyDQ}&\rotatebox{90}{OpenRefine \& MetricDoc~}&\rotatebox{90}{Oracle EDQ}&\rotatebox{90}{SAS Data Quality}&\rotatebox{90}{Talend Open Studio}\\\hline\hline
         39&Task scheduling&$p$&\checkmark&$-$&\checkmark&$p$&\checkmark&\checkmark&$p$&\checkmark&$-$&\checkmark&$p$&$-$\\
         40&Storage of results&\checkmark&\checkmark&\checkmark&\checkmark&\checkmark&\checkmark&\checkmark&\checkmark&\checkmark&\checkmark&\checkmark&\checkmark&\checkmark\\
         41&Retrieval of results&$-$&\checkmark&\checkmark&\checkmark&$-$&$-$&\checkmark&\checkmark&$-$&$-$&\checkmark&\checkmark&\checkmark\\
         42&Comparison&$-$&\checkmark&$-$&\checkmark&$-$&\checkmark&\checkmark&\checkmark&$-$&$-$&$-$&$p$&$-$\\
         43&Visualization over time&$-$&\checkmark&$-$&\checkmark&$-$&\checkmark&\checkmark&\checkmark&$-$&$-$&$-$&$p$&$-$\\\hline
\end{tabular}
\end{center}
\end{table}

%% Storage (CDQM-40) + retrieval (CDQM-41)
The storage (CDQM-40) of DP or DQM results is possible in all tools. The majority of DQ tools (except MobyDQ and OpenRefine) also support data export via a GUI. Datamartist allows to export only very basic data profiles. 
The most comprehensive enterprise solutions for CDQM-40 and 41 is provided by Informatica DQ and SAS Data Quality, which enable the export of full DP procedures. During import, all settings and required data sources are reloaded from the time of the analysis. 

%% Task scheduling, comparison, visualization
Task scheduling (CDQM-39) is also widely supported. Aggregate Profiler fulfills this requirement only partially, since it is only possible to schedule business rules, but no other form of tasks, e.g., data profiling tasks. With Datamartist, InfoZoom, and SAS Data Quality, task scheduling is cumbersome for business users, since the command line is required to write a batch file. 
In the case of Datamartist, the tool needs to be closed in order to allow the execution of the batch file.

To visualize the continuously performed DQ checks (be it DP tasks, user-defined rules, or DQ metrics), Informatica DQ relies on so called ``scorecards'', which can be customized to display the respective information. 
Apache Griffin, Experian Pandora, and SAS Data Quality also allow alerts to be defined, when specific errors occur or when a defined rule is violated. MobyDQ does not offer any visualization (which is considered future work), but  relies on external libraries in its implementation at Ubisoft.
SAS fulfills both requirements CDQM-42 and 43 only partially, since its ``dashboards'' contain solely the number or percentage of triggers per date, source or user, but no specific values (e.g., 80~\% completeness) could be plotted. 
The most comprehensive solution for CDQM in general-purpose DQ tools provide Informatica DQ and DataCleaner by Human Inference. With respect to the open-source tools, only Apache Griffin provides comprehensive CDQM support, and the commercial version of MobyDQ, which is deployed at Ubisoft.

\section{Survey Discussion and Lessons Learned}
\label{sec:survDisc}
% Results and Discussion
% Strengths and Weaknesses
% Meaning of findings
The results of our survey on DQ measurement and monitoring tools revealed interesting characteristics of DQ tools and allow to draw conclusions about the future direction of automated and continuous data quality measurement. 

\subsection{Overview on the Market of DQ Tools}
One of the greatest challenges we faced during the conduct of this survey was the constant change and development of the DQ tools, especially the open-source tools. Nevertheless, it is of great value to reflect on the current state of the market for two reasons: (1) to create a uniform vision for the future of DQ research, and (2) to identify the potential for functional enhancement across the tools. 

The fact that we found 667 tools attributed to ``data quality'' in our systematic search indicates the growing awareness of the topic. However, approximately half (50.82~\%) of the DQ tools that we found were domain specific, which means they were either dedicated to specific types of data or built to measure the DQ of a proprietary tool. This amount underlines the ``fitness for use'' principle of DQ, which states that the quality of data is dictated by the user and type of usage. 40~\% of the DQ tools were dedicated to a specific data management task, for example, data cleansing, data integration, or data visualization, which reflects the complexity of the topic ``data quality''. Although  those tasks are often not clearly distinguished in practice, we required explicit DQ measurement, that is, making statements about the DQ without modifying the observed data. 

Our selection of DQ tools provides a good digest of the market, since we included nine commercial and closed-source tools as well as five free and open-source tools, from which four (except Talend) are not mentioned by Gartner~\cite{GartnerDQTools2019}. 
The vendors of four tools have been named ``leader'' in the Magic Quadrant of Data Quality Tools 2019 (Informatica, SAS, Talend, Oracle) and two of them are among the four vendors currently controlling the market (which are SAP, Informatica, Experian, and Syncsort~\cite{GartnerDQTools2019}; however no trial for SAP nor Syncsort was granted). 
Overall and according to our requirements, we experienced Informatica DQ as the most mature DQ tool. 
The best support for data profiling is provided by Experian Pandora, which allows to profile across an entire DB and even across multiple connected data sources. All other tools allow data profiling only for selected columns or within specific tables. Despite being classified as leader by Gartner, we perceived Oracle EDQ, Talend OS, and SAS Data Quality as having less support for data profiling and/or DQ monitoring. Although Quadient (with DataCleaner) was removed from the Gartner study in 2019 due to their focus on customer data, our evaluation yielded a good support in data profiling and a strong support in DQ monitoring. 
However, when comparing the two general-purpose and freely available DQ tools Talend OS and Aggregate Profiler, the former one convinced in terms of intuitive user interface and a good overall performance.
Aggregate Profiler on the other hand, has a richer support for advanced multi-column profiling and data cleansing, but it is not always clear which algorithms are used to perform data modifications and the documentation is not up-to-date.
Three open-source tools (Apache Griffin, MobyDQ, and OpenRefine) were installed from GitHub and thus required technical knowledge for the setup. While OpenRefine can not keep up with comparable tools like Talend OS or Aggregate Profiler in terms of data profiling, MobyDQ and Apache Griffin have clearly a different focus on CDQM. 
IBM ISDQ demonstrated, that also commercial tools can be very arduous and time intensive to install due to the increasing complexity of the single modules and  dependencies between them.

\subsection{Data Profiling and its Relevance for Data Quality}
The requirements in our data profiling category are based on the classification by Abedjan et al.~\cite{Abedjan2019}, who point out that there is no clear distinction between data profiling and data mining. According to~\cite{Abedjan2019}, the two topics can be distinguished by the object of analysis (focus on \textit{columns} in data profiling vs. \textit{rows} in data mining) and by the goal of the task (gathering technical metadata by data profiling vs. gathering new insights by data mining). While this distinction is still fuzzy, we go one step further and claim that there is also no clear distinction between data mining and data analytics with respect to the used techniques (e.g., regression analysis is discussed in both topics~\cite{DasuJohnson_2003}). 
Our evaluation revealed that single column profiling (DP 1-18) and dependency discovery (DP 19-24) are typically implemented within data profiling views in the DQ tools, despite a partial overlap of the techniques with the data mining domain (e.g., histograms and quartiles are discussed in both research areas).
However, DQ tools rarely support advanced multi-column profiling requirements (DP 25-30), like clustering or association-rule mining. According to our customer contacts and reference customers, those functionalities are not considered to be part of data profiling and are usually implemented in analytics tools (e.g., SAS Enterprise Guide). 
This might be a reason why most DQ tools in our evaluation lack a wide range of features in this category: customers and vendors simply do not consider it as part of data profiling and data quality. 

In recent years, numerous research initiatives concerning data profiling have been carried out that also use ML-based methods. Current general-purpose DQ tools do not take full advantage of these features.
Although, several vendors claim to implement ML-based methods, we found no or only limited documentation of concrete algorithms (cf.~\cite{Quadient_doku}). 
Note that in the case of DataCleaner for duplicate detection, we received more detailed documentation upon request. 
We think that especially concerning the hype for artificial intelligence and the enhancement of detecting DQ errors with ML methods, it is necessary to focus on the desirable core characteristics for DQ and data mining~\cite{DasuJohnson_2003}: the methods should be widely applicable, easy to use, interpret, store and deploy, and should have short response times. 
A counterexample are neural networks, which are increasingly applied in recent research initiatives, but need to be handled with care for DQ measurement, because they are black-box and hard to interpret. 
For measuring the quality of data (to ensure reliable and trustworthy data analysis), easy and clearly interpretable statistics and algorithms are required to prevent a user from deriving wrong conclusions from the results. 

\subsection{Data Quality Monitoring}
One of the main objectives of this survey was to investigate the CDQM features in DQ tools, as there are no scientific evaluations and statements on this topic.
In contrast to Pushkarev et al.~\cite{Pushkarev_DQSurvey_2010}, who did not find any tool that supports DQ monitoring, we clearly identified the existence of this feature, as shown in Table~\ref{tab:survey_cdqm}. 

In general-purpose DQ tools, DQ monitoring is considered a premium feature, which is liable to costs and only provided in professional versions. An exceptions is the dedicated open-source DQ monitoring tool Apache Griffin, which supports the automation of DQ metrics, but lacks pre-defined functions and data profiling capabilities.
The remaining open question with respect to DQ monitoring is which aspects of the data should actually be measured (discussed in the following paragraph). 
An additional challenging topic for future research is automated analysis of the CDQM time-series  in order to predict trends and sudden changes in the DQ~\cite{EhrlingerWoess_2017}.

\subsection{How to Measure Data Quality}
\label{sec:survDisc_mes}
In our survey, we did not find a tool that implements a wider range of DQ metrics for the most important DQ dimensions as proposed in research papers (cf.~\cite{Heinrich_2018,Piro_2014,BatiniScannapieco_2016}) and we also did not find another survey that investigates the existence of DQ metrics in tools. 
Identified DQ metric implementations have several drawbacks: some are only applicable on attribute-level (e.g., no aggregation possibility), some require a gold standard that might not exist, and some have implementation errors. 
%% Accuracy and the problem of gold standard
The two open-source tools that implement metrics for the DQ dimensions accuracy (Apache Griffin) and completeness between two tables (MobyDQ) relied on a reference data set (i.e., gold standard) provided by the user. Apache Griffin based their metric on the definition by DAMA UK, who state that accuracy is ``the degree to which data correctly describes the `real world' object or event being described''~\cite{DAMA_dimensions}, which needs to be selected for the calculation. 
MobyDQ specifically aims at automating DQ checks in data pipelines, that is, computing the difference between a source and a target data source, where the gold standard is clearly defined. However, in scenarios where the quality of a single data source should be assessed, such metrics are not suitable since a reference or gold standard is often not available~\cite{QuaIIeJournal_2018}. This fact is also reflected by the restricted prevalence of such gold-standard-depending DQ metrics in commercial and general-purpose DQ tools. 

The other investigated tools mainly implement two very basic metrics: completeness (indicating the missing data problem) and uniqueness (indicating duplicate data values or records). 
%% Completeness + Uniqueness
It is noteworthy that while completeness is one of the most-widely used DQ dimensions (cf.~\cite{CCDQ,Heinrich_2018,BatiniScannapieco_2016}), the aspect of uniqueness is often neglected in DQ research~\cite{Ehrlinger_Minimality}. For example, Piro~\cite{Piro_2014} perceive duplicate detection as a \textit{symptom} of data quality, but not as DQ dimension. 
Neither Myers~\cite{CCDQ} in his ``List of Conformed Dimensions of Data Quality'', nor the ISO 25012 standard on DQ~\cite{ISO25012} refer to a DQ dimension that describes the aspect of uniqueness or non-redundancy~\cite{Ehrlinger_Minimality}. Despite this difference, both DQ dimensions have a common characteristic: they can be calculated without necessarily requiring a gold standard.
%% What is missing: aggregation; instance-level vs. schema-level dimensions 
Nevertheless, these implementations lack two aspects: (1) the aggregation of DQ dimensions and (2) schema-level DQ dimensions that are clearly part of the DQ topic~\cite{BatiniScannapieco_2016}. 
The aggregation of DQ dimensions from value-level to attribute-, record-, table-, DB- or cross-data-source-level as presented in~\cite{Hinrichs_2002,Piro_2014} was not provided  by any tool prefabricated. Informatica DQ is the only tool that allows to aggregate column-level metrics on table-level, but not higher.
We did not declare a manual implementation in tools with strong rule support as availability of such aggregation functions. 

%% Methodologies, David Loshin, DAMA UK in Griffin
Despite the lack of prefabricated DQ metrics, most tools refer to a set of DQ dimensions in their user guide or defined methodology, for example, Informatica and SAS rely on whitepapers influenced by David Loshin~\cite{SAS_doku, Informatica_DQ_methodology}, or Talend promotes the existence of such metrics on their website\footnote{\url{https://info.talend.com/vrstosdq_150602.html} (June 2019)}.
In Section~\ref{sec:eval_DQM}, we showed that the list of referenced DQ dimensions and metrics by the DQ vendors is very non-uniformly. Further inquiry on the metrics yielded two different responses by our customer contacts: while some explicitly stated that they do not offer generally applicable DQ metrics, others could not answer the question of how specific metrics are implemented. 

In the case of Talend, we asked our customer contact and the Talend Community\footnote{\url{https://community.talend.com} (June 2019)}, where the metrics promoted on the website can be found. Unfortunately, we got no satisfying answer, only references to the data profiling perspective in TOS and its documentation. This experience underlines the statement by Sebastian-Coleman~\cite{Coleman_2012}, that ``people can often not say how to measure completeness or accuracy'', which also leads to the different interpretations and implementations.

Other vendors justified the absence of generally applicable DQ metrics with two reasons: because such metrics are not feasible in practice, and because customers do not request it. Several DQ strategies also indicate the fact that DQ metrics should be created by the user and adjusted to the data (cf.~ \cite{Griffin_doku,Informatica_DQ_methodology,SAS_doku}). This understanding follows the ``fitness for use'' principle, which highlights the subjectivity of DQ. 
Also Piro~\cite{Piro_2014} states that objectively measurable DQ dimensions previously require a manual configuration by a user.
An example for this is Apache Griffin, who state that ``Data scientists/analyst \textit{define} their DQ requirements such as accuracy, completeness, timeliness, and profiling''~\cite{Griffin_doku}. 
Sebastian-Coleman points out that it is important to understand the DQ dimensions, but these do not immediately lend themselves to enabling specific measurements~\cite{Coleman_2012}. 
The main foundation into DQ measurement, including the set of DQ dimensions and metrics have been originally proposed in the course of the Total Data Quality Management (TDQM) program of MIT\footnote{\url{http://web.mit.edu/tdqm} (June 2019)} in the 1980s. 
In 2003, Dasu and Johnson~\cite{DasuJohnson_2003} state that DQ dimensions, as originally proposed by the TDQM, are not practically implementable and it is often not clear what they mean. The results of our survey underlines this statement with a scientific foundation, because each DQ tool implements the dimensions differently, and partially far away from the complex metrics proposed in research (e.g., no aggregation, often no gold standard).
Apart from completeness and uniqueness on attribute-level, no DQ dimension finds wide-spread agreement in the implementation and definition in practice. This is especially noteworthy for the frequently mentioned accuracy dimension, which however, requires a reference data set that is often not available in practice. 

%% Consequence: question metrics + dimensions
%% Conclusion, we don't really need the dimensions?
We conclude that there is a strong need to question the current usage of DQ dimensions and metrics. 
Research efforts to measure DQ dimensions directly with a single, generally-applicable DQ metric have little practical relevance and can hardly  be found in DQ tools. 
In practice, DQ dimensions are used to group domain-specific DQ rules (sometimes referred to as metrics) on a higher level. 
Since research and practitioners failed to create a common understanding of DQ dimensions and their measurement for decades, a complementary and more practice oriented approach should be developed. 
Several DQ tools show that DQ measurement is possible without referring to the dimensions at all. 
Since our focus is the automation of DQ measurement, a practical approach would be required without the need for DQ dimensions, but a focus on the core aspects (like missing data and duplicate detection), which can actually be measured automatically. 

\subsection{The Requirement for Automation and Declaration}
Current DQ tools provide a wide range of different data profiling features and allow the user to define rules to which the data must conform. In order to be able to use such a DQ tool in practice, it is initially necessary to manually create and adjust the DQ measurements. 

According to reference customers, more automated, initial and still meaningful out-of-the-box profiling across tables and data sources would be required, for example, to highlight abnormal data values. 
At the same time, it is necessary to bridge the gap between prefabricated DQ measures and insight into the performed calculations and algorithms. In several tools (e.g., AggregateProfiler, InfoZoom), plots were generated or outliers were detected without a clear declaration of the used threshold or distance function. 
In alignment with the requirement for interpretability of data profiling results, we highlight the need for clear declaration of the parameters used. 

\section{Conclusion and Outlook}
\label{sec:conclusion}
% summary of what we did
In this survey, we conducted a systematic search in which we identified 667 software tools dedicated to the topic ``data quality''. With six predefined exclusion criteria, we extracted 17 tools for deeper investigation. We evaluated 13 of the 17 tools with regard to our catalog of 43 requirements  divided into the three categories (1) data profiling, (2) DQ measurement, and (3) continuous DQ monitoring. Although the market of DQ tools is continously changing, this survey gives a comprehensive overview on state-of-the-art of DQ tools and how DQ measurement is currently perceived in practice by companies in contrast to DQ research.

% summary of findings
So far, there are only a few surveys on DQ tools in general, and in particular no surveys that investigated the existence of generic DQ metrics. There is also no survey that could identify the existence of DQ monitoring capabilities in DQ tools. 
We close this gap with our survey and provide the results regarding the available DQ metrics and DQ monitoring capabilities for the tools analyzed.

% summary of future work
In our ongoing and future work, we aim to introduce a practical DQ methodology that regards at directly measurable aspects of DQ in contrast to abstract dimensions with no common understanding. 
We also think that it is worth investigating the potential for automated out-of-the-box data profiling along with a clear declaration of the used parameters, which might be modified after the initial run. 
Part of our ongoing research is to exploit time-series analytics for further investigation of DQ monitoring results. 
Since a deep investigation of single DP requirements was out of scope for this survey, it would also be worth to further investigate specific implementations and their proper functionality, for example, which aspects yield floating point differences. 

The top vendors of DQ tools worldwide have between 7,200 (Experian), 5,000 (Informatica) and 2,700 (SAS) customers for their DQ product line~\cite{GartnerDQTools2019}. Compared to the hype for AI and ML, these low numbers show high catch-up demand for DQ tool applications in general. 
\appendix
\section{Appendix: Data Profiling Test Cases}
\label{apx:db_schema}
For the evaluation of data profiling requirements, we employed a modernized version of the well-known Northwind DB published by dofactory\footnote{\url{http://www.dofactory.com/sql/sample-database} (June 2019)}. 
The following list comprises all test cases we performed, where the enumeration can be linked to the DP requirements from Table~\ref{tab:reqCatalog}:
 \begin{enumerate}
 	\item Number of rows in table \texttt{Product}
 	\item Number of \texttt{null} values in column \texttt{Supplier.Fax}
 	\item Percentage of \texttt{null} values in column \texttt{Supplier.Fax}
 	\item Number of distinct values in column \texttt{Customer.Country}
 	\item Number of distinct values divided through number of rows for \texttt{Customer.Country}
 	\item Frequency histograms for \texttt{Customer.Country}
 	\item Minimum and maximum values in \texttt{OrderItem.UnitPrice}
 	\item Constancy for column \texttt{Customer.Country}
 	\item Quartiles in column \texttt{OrderItem.UnitPrice}
 	\item Distribution if first digit=1 in column \texttt{UnitPrice}, table \texttt{OrderItem}
 	\item Basic types for \texttt{ProductName}, \texttt{UnitPrice}, and \texttt{isDiscontinued} in table \texttt{Product}
	\item DBMS-specific data types for \texttt{ProductName}, \texttt{UnitPrice}, and \texttt{isDiscontinued} in table \texttt{Product}
	\item Minimum, maximum, average, and median value length of column \texttt{Product.ProductName}
 	\item Maximum number of digits in column \texttt{Product.UnitPrice}
 	\item Maximum number of decimals in column \texttt{Product.UnitPrice}
 	\item Count of pattern ``AA'' in \texttt{Customer.Country}, derived from histogram
 	\item Semantic data types for \texttt{ProductName}, \texttt{UnitPrice}, and \texttt{isDiscontinued} in table \texttt{Product}
	\item Semantic domains for \texttt{ProductName}, \texttt{UnitPrice}, and \texttt{isDiscontinued} in table \texttt{Product}
 	\item All 100 \% conforming UCCs in \texttt{Order}
 	\item All 98 \% conforming UCCs in \texttt{Order}
 	\item All 100 \% conforming INDs between \texttt{Order.CustomerId} and \texttt{Customer.Id}
 	\item All 93 \% conforming INDs between \texttt{Order.CustomerId} and \texttt{Customer.Id}
 	\item All 100 \% conforming FDs in \texttt{Order}
 	\item All 93 \% conforming FDs in \texttt{Order}
 	\item Correlation between \texttt{OrderItem.UnitPrice} and \texttt{OrderItem.Quantity}
 	\item All possible association rules within \texttt{Product}.
 	\item Clustering the values in  \texttt{Product.UnitPrices}
 	\item All ``very high values'' in \texttt{Order.TotalAmount}
 	\item All exact duplicates in \texttt{Customer}, considering \texttt{FirstName} and \texttt{LastName} only
 	\item All relaxed duplicates in \texttt{Customer}, considering \texttt{FirstName} and \texttt{LastName} only
\end{enumerate}

\section*{Acknowledgement}
The research reported in this paper has been supported by the Austrian Ministry for Transport, Innovation and Technology, the Federal Ministry for Digital and Economic Affairs, and the Province of Upper Austria in the frame of the COMET center SCCH. 

The authors would like to thank all contact persons who provided us with trial licences and support, in particular, Thomas Bodenm{\"u}ller-Dodek and Dagmar Hillmeister-M{\"u}ller from Informatica, David Zydron from Experian, Alexis Rolland from Ubisoft,  Marc Kliffen from Human Inference, Ingo Lenzen from InfoZoom, Loredana Locci from SAS, and Rudolf Plank from solvistas GmbH. Special thanks to Alexander Gindlhumer for his support with Apache Griffin. 

\bibliographystyle{plain}  
\bibliography{dqtoolsurvey}  %%% Remove comment to use the external 
\end{document}